\documentclass{article}

\usepackage{arxiv}

\usepackage[utf8]{inputenc} 
\usepackage[T1]{fontenc}    
\usepackage[hidelinks]{hyperref}      
\usepackage{url}            
\usepackage{booktabs}       
\usepackage{amsfonts}       
\usepackage{nicefrac}       
\usepackage{microtype}      
\usepackage{lipsum}		
\usepackage{graphicx}
\usepackage{natbib}
\usepackage{doi}
\usepackage{rotating}
\usepackage{subfig}
\usepackage{makecell}
\usepackage{xcolor}
\usepackage[normalem]{ulem}
\usepackage{soul}
\usepackage{ulem}
\usepackage{comment}

\usepackage{floatrow}
\usepackage{caption}
\usepackage{subcaption}
\usepackage[rightcaption]{sidecap}

\title{Is the panel fair? Evaluating panel compositions through network analysis. The case of research assessments in Italy}

\date{} 					

\author{\href{https://orcid.org/0000-0003-0293-482X}{\includegraphics[scale=0.06]{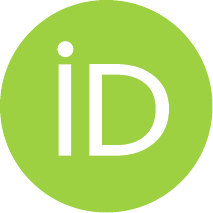}\hspace{1mm}Alberto Baccini}
  \\
	Dipartimento di Economia Politica e Statistica\\
	Università degli Studi di Siena\\
	Siena, Italy \\
	\texttt{alberto.baccini@unisi.it} \\
	\And
	\href{https://orcid.org/0000-0001-7715-8250}{\includegraphics[scale=0.06]{orcid.pdf}\hspace{1mm}Cristina Re} \\
	Dipartimento di Economia Politica e Statistica\\
	Università degli Studi di Siena\\
	Siena, Italy \\
	\texttt{cristina.re@unipr.it} \\
}

\renewcommand{\headeright}{}
\renewcommand{\undertitle}{}
\renewcommand{\shorttitle}{Is the panel fair?}

\hypersetup{
pdftitle={Is the panel fair?},
}

\makeatletter
\input{size11.clo}
\makeatother

\begin{document}

\maketitle

\begin{abstract}
Research evaluation is usually governed by panels of peers. Procedural fairness refers to the principles that ensures decisions are made through a fair and
transparent process. It requires that the composition of panels is fair. A fair panel is usually defined in terms of observable characteristics of scholars such as gender or affiliations. The formal adherence to these criteria is not sufficient to guarantee a fair composition in terms of scholarly thinking, background, or policy orientation. 
An empirical strategy for exploring the fairness in the intellectual composition
of panels is proposed, based on the observation of links between panellists.
The case study regards the three panels selected to evaluate research in economics, statistics and business during the Italian research assessment exercises. The first two panels were appointed directly by the governmental agency responsible for the evaluation, while the third was randomly selected. Hence the third panel can be considered as a control for evaluating about the fairness of the others. The fair representation is explored by comparing the networks of panellists based on their co-authorship relations, the networks based on journals in which they published and the networks based on their affiliated institutions (universities, research centres and newspapers). The results show that the members of the first two panels had connections much higher than the members of the control group. Hence the composition of the first two panels should be considered as unfair, as the results of the research assessments. 

\end{abstract}


\newpage

\section{Introduction}
Research evaluation has now become one of main instruments for university and research policies. It has been “integrated as administrative routine at many
levels and in many types of institutions” \citep[p. 2]{dahler-larsen}. It is no longer considered as just an essential step for distributing resources to competing research projects and for hiring and promoting researchers; it is also the core of the administrative processes labelled ``performance based research funding'' where ex-post evaluation is used for financing universities and other research institutions \citep{langfeldt2004,Hicks,Zacharewicz}. The consequences are twofold. First, research evaluation has gained centrality in the scientific knowledge-making process, so much that it has taken ``the function of gatekeeping, filtering, and legitimating'' it \cite[p. 209]{lamont2011}. Second, research evaluation, being entered in administrative processes, should follow their rules, and among these the ones regarding procedural fairness \citep{Ruder_woods}. 

Procedural fairness is ``concerned with procedures used to arrive at {[}fair{]} outcomes'' \cite[p. 220]{beersma2003}, i.e. procedures should be ``designed to reduce bias or favoritism in government decision making'' \citep[p. 400]{Ruder_woods}. In general, procedural fairness aims to protect individuals' rights and maintain trust in the integrity and fairness of administrative decision-making. Procedural fairness in an administrative procedure refers to the principles that ensures decisions are made through a fair and transparent process. It typically involves the right to an unbiased decision-maker: decisions must be made by an impartial authority who does not have any personal interest, bias, or preconceptions that could influence the outcome \citep{Leventhal1980, bobocel}.

Despite their concrete variety, research evaluation is usually governed by panels of experts or peers that managed or directly realized the processes of peer review \citep{whitley2010}. Hence, the composition of these panels is a key issue for procedural fairness, which generally requires \emph{fair representation} of all affected parties involved in the decision-making process \citep{Leventhal1980}. The problem is how to define who the affected parties are and what dimensions to take into account in defining their fair representation. The usual way in which this problem is handled is by considering observable non-scientific characteristics of scholars, such as gender, age, affiliation, geography that may influence evaluation practices. Also other scientific characteristics of scholars are considered, such as their field of research or scientific prominence in order to ensure their ability to make an expert evaluation. The building of fair panels requires a proper balance of all these characteristics.
The European Peer Review Guide of the European Science Foundation \citep{esf2011}, the rules for the composition of panels of British Research Excellence Framework (REF) and of the Italian research assessments are examples of the way in which institutions concretely try to build fair evaluation panels. 

An unbalanced panel composition is usually seen as a real risk that certain groups will capture the evaluation systems. Many studies have examined how panel composition affects evaluation outcomes and the resulting distribution of awards. They analyzed evaluation procedures in view of showing the respect or the departure from the Mertonian universalism \citep{Merton}, according to which the evaluation and the distribution of awards should reflect the distribution of ``intellectual deservingness of recipients, rather than to their group identity or status or their particularistic relationship with the panellists'' \citep[575]{Mallard}. In doing so, these studies focused on the relationship between the distribution of awards or results of evaluation and the social characteristics (e.g., age, race, gender, scientific prominence) of panellists or of individuals whose work is evaluated. They regarded mainly cases of \emph{ex-ante research evaluation} \citep{cole1992, cole1981, cole1979, lamont2011, Rahman}, but also cases of \emph{ex-post research evaluation,} such as the national research assessments in United Kingdom and Italy \citep{lee2007, lee2013, harley1997, baccini2016, Baccini_De_Nicolao_2021, corsi2010, corsi2011}. 

The research about ``procedural fairness'' in research evaluation has considered also the processes through which panels arrive at final decisions \citep{Mallard}. The specific composition of a panel contributes to the definition of the ``set of informal, and sometimes inconsistent, procedural rules that the panellists tacitly co-construct and then adhere to in the course of their deliberations'' \citep[p. 18]{Camic}. Hence, the way panels are set up may generate cronyism, the pursuit of self-interest, and cognitive particularism in peer review \citep{lamont2011}. In this literature, the relevant dimension in the composition of a panel is disciplinary, because scientists are socialized in evaluative practices which are discipline specific. These practices are the basis from which panel members interact to reach evaluative deliberations \citep{lamont2011}. In sum, a fair composition of a panel in terms of observable characteristics of its members is not sufficient to guarantee that the composition of the panel is also, so to speak, ``intellectually'' fair. The lack of a fair intellectual composition can be considered as a way in which procedural unfairness emerges in research evaluation. Similarly, when a panel is called to evaluate a discipline characterized by the coexistence of many schools of thought with different approaches, methodologies, policy recipes and evaluative practices, its sub-disciplinary composition may be relevant to its deliberations. In this case an intellectual unfair selection of panellists may contribute to the prevalence of one of competing paradigms or schools of thought, by reinforcing the normative standard, pre-existing journal rankings and more generally the  hierarchy of a discipline. 

Probably, economics is the most studied case  of the dysfunctional effects \citep{ferguson2018} of unfair compositions of disciplinary panels in research evaluation (see \cite{corsi2019} for a review of the literature). Pioneer of these studies was Frederic S. Lee, who analyzed the Research Assessments Exercise (RAE) in the United Kingdom \citep{harley1997, lee2007, lee2013}. He explicitly individuated a precise role for the panel of RAE, composed by  paradigmatic homogeneous experts ``controlled by mainstream economists [who] have used [the RAE] to support particular neoclassical research over heterodox research and promote neoclassical departments over more pluralistic ones'' \cite[p.15]{lee2007}. A process aimed ``to achieve a discipline-desired outcome that was (and is) compatible with the Government's pro-market ideological agenda'' \cite[p.14]{lee2007}. The Italian case also has been considered as an internationally relevant example on how it can disregard heterodox schools and historical methods in favour of mainstream approaches and quantitative methods in economics \citep{Pasinetti,corsi2010, corsi2011}, probably connected to  a cultural and political change from the Keynesian to the Ordoliberal ideology \citep{re2019}. The lack of fairness in the composition of the committees in charge of economic evaluation was also highlighted: their homogeneity could probably have minimized the voices of disagreement with the evaluation methods and rules adopted \citep{Pasinetti, BACCINI2011, baccini2014, baccini2016, BACCINIricc2012}. 

This paper proposes an empirical strategy for reasoning about the fairness in the intellectual composition of panels. It attempts to detect connections among members such that panel composition can be considered as unfair, despite formal adherence to a fair panel composition in terms of easily observable characteristics such as gender, affiliation or geography. The social and intellectual connections between panel members are investigated through network analysis techniques. Three kinds of connections, and hence three kinds of networks are considered and explored: (i) the coauthorship networks built by considering the publications of the members of the panels; (ii) the journal based networks built by considering the journals where the panellists publish their scholarly articles; and (iii) the ``affinity networks'' built by considering panellists' affiliations to universities and research centres, and their collaborations to newspapers or blogs. 

The main difficulty of this kind of approach is to define a threshold of connections above which the panel composition should be considered unfair. The basic intuition is that the panel should be considered unfair if the connections between its members are anomalously higher than the normal connections that occur in the population of scholars from which the panel was chosen. Unfortunately, this threshold can be difficult, if not impossible, to calculate in practice, if the population of scholars is too large or not clearly defined. The case study developed here regards the composition of the panels appointed to evaluate research in economics, statistics and business during the Italian research assessment exercises, the so called VQR 2004-2010, VQR 2011-2014 and VQR 2015-2019. The three panels were selected by ANVUR, the governmental agency for the evaluation of university and research, by using different rules. Indeed, as it is detailed later on, the first two panels were appointed directly by the governing board of ANVUR; the third panel was instead selected randomly by a lot. Hence this third panel can be considered as a sort of control group: the connections between members of  the third panel can be considered as the reference threshold for judging the fairness of the composition of the other two panels. 

The organization of the article is as follows: Section 2 reviews literature on procedural fairness and panel composition; in Section 3, the Italian research assessments and the panel selection methods are described; in Section 4 the research design is presented; sections 5-7 report the analysis, respectively, of the co-authorship networks, of the journal based networks, and of the affinity networks. Section 8 discusses results and concludes.

\section{Research evaluation, procedural fairness and panel composition}

In a literature review, \citet{Mallard} stated that studies of peer review and research evaluation had until then largely focused on \emph{distributive fairness} by restricting the analysis only to the final results of evaluation and the consequent distribution of rewards or punishments. Indeed, the so called ``first wave'' of research on peer review and research evaluation \citep{Lamont_2005} was mainly focused on obstacles to “fairness caused by nonscientific influences such as politics, friendship networks, or common institutional positions” \citep{travis1991}. Both potential and observed risks in the peer review system were discussed and documented. It was argued that the system is conservative and suppresses innovative research. Several scholars even suggested that peer review hinders scientific progress \citep{horrobin1990}.  Effects such as nepotism and old-boyism in peer review were seen to hinder pioneering
research \citep{chubin1990, roy1985}, while ``cognitive
particularism'', ``favoritism for the familiar'' and ``scholarly bias''
support the school viewpoint or research topic the reviewers themselves
are conducting (see, e.g., \cite{porter1985, travis1991}). Moreover, another possible group effect is groupthink, which refers to ``a deterioration of mental efficiency,
reality testing, and moral judgement that results from in-group
pressures'' \cite[p.9]{janis1982}. Loyalty to the group ``requires each
member to avoid raising controversial issues, questioning weak
arguments, or calling a halt to soft-headed thinking'' \cite[p.12]{janis1982}. 

The empirical literature of this first wave of research largely that universalistic norms were followed more often than not \citep{gao1994, cole1979, cole1981, Zuckerman_merton}. Subsequent studies produced mixed evidence. \citet{wanneras1997} argued that peer review fosters nepotism and gender bias, disadvantaging women, though subsequent studies have not consistently supported these claims \citep{marsh2009}. Other studies suggest that gender dynamics in academic evaluations may indeed play a role: \citet{van2010} found that in the Netherlands, an increased presence of women on appointment committees led to more women being appointed as full professors, suggesting a preference for same-gender candidates. \citet{langfeldt2004} documented a variety of source of bias in panels appointed to evaluate research. These biases emerged from interactions (discussion, negotiations, lack of agreement) between panel members, for which the selection of panel was crucial.

According to \citet{Mallard}, the first wave of studies did not systematically consider the question of ``procedural fairness'', i.e. the fairness of the procedures used in research evaluation to produce evaluative results. That question started to be explored by a second wave of literature on research evaluation and peer review, focusing in particular on cognitive dimensions of peer evaluation. \citet{Collins_evans} discussed the nature and role of scientific expertise in decision making processes. \citet{Lamont_2005,Mallard,lamont2011} proposed a very specific interpretation of procedural fairness, as the way in which panels of evaluators concretely construct the evaluation process, i.e. how they define and apply rules and criteria of evaluation. \citet{lamont2011} analyzed the customary rules constructed by some panellists in research evaluation and how they perceive the outcome of their decisions as ``fair''. Panellists tend to perceive their decision as fair in reference to the concrete rules and process they used for handling disagreement and negotiating agreement. panellists consider fair an evaluative procedure if they shared the ``belief that meritocracy guides the process, while corrupting forces, self-interest, and, in particular, politics are kept at bay'' \citep[p. 214]{lamont2011}. In summary, according to this literature, research evaluation is shaped and constrained by the procedural context in which occurs. It is therefore useless to contrast biased and unbiased evaluation, because ``extra-cognitive factors do not corrupt the evaluation process but are intrinsic to it'' \citep[p. 228]{lamont2011}. In this regard, the structure and composition of panels is of primary interest because it can influence the ``customary rules'' concretely adopted in research evaluation. An ``unfair evaluation'' may result from a panel composition in which evaluators agree to adopt rules that favor research similar to their ones in terms of methodological or epistemological features, research styles or policy prescriptions.

The focus on procedural fairness can be understood in reference also to different streams of literature regarding the institutional frame of research evaluation and peer review. The first step of the reasoning consists in considering that research evaluation and peer review are integrated as administrative routine in many types of institutions \citep{dahler-larsen}. Indeed, research evaluation is carried out by governments, academic institutions, agencies,  or other organizations to produce indicators or other evaluation information. 
\citet[p. 6]{pievatolo_2024} highlighted that in the administrative evaluation of research, ``assessment authorities'' give to evaluators decision-making power. When scholars are in charge of evaluation in an administrative procedure, they are subtracted to the public scrutiny of their opinion as usually happen in the scientific debate, because they can impose their decisions ``even on those who disagree''\citep [p. 6] {pievatolo_2024}. Hence, if research evaluation has fully entered into administrative activities such as recruitment, selection, funding, it will eventually follow its rules. Procedural fairness - sometimes also referred to as procedural justice or procedural due process - is a cornerstone of administrative processes. They are fair when they are designed to reduce bias or favoritism in decision making. Fairness may be judged in terms of a procedure's ``consistency over time and across persons; its accuracy and prevention of personal bias; or its representativeness of the values, interests, and outlook of important subgroups in the population of persons affected'' \cite[p.54]{Leventhal1980}. Transparency and public participation are elements of procedural fairness. It is the main source of legitimacy of administrative decisions made by government agency \citep{Ruder_woods}. It is a signal that people subjected to the decision are respected, that decision-makers can be trusted and that the final decision can be considered fair \citep[p. 293]{Esaiasson}. According to \citet{Frey}, procedural fairness generates procedural utility because rational people have preferences about how outcomes are generated.

If we consider research evaluation as a special kind of administrative procedure, its fairness concerns also the composition of evaluation panels. \citet{BACCINIricc2012} suggested an analogy between the role of panellists in research evaluation and that of members of a popular jury in a trial. In order to have a fair legal judgment by a panel of judges, it is necessary to designate a fair jury and therefore presumably less inclined to partiality.  Similarly, in order to have a fair research evaluation, the panel should be composed in a fair manner. 

The fair procedures to select a jury is a widely debated issue. Randomization is considered as an effective way to build a fair jury, by giving a fair chance to every qualified citizen to serve on a jury, and to provide defendants and litigants to be tried by a ``representative cross-section of the population'' \citep[p. 75]{duxbury}. The rules of the United States \emph{Jury Selection Service Act}, for example, state that the jury must be
appointed by selecting ``at random from a fair cross-section of the
community''. The fair composition of the popular jury aims to ensure
fairness of the judgment. ``Achieving representative cross-sections of
the community in jury venires, and ensuring that our civil juries
reflect the community as well, are essential components contributing to
the fairness and legitimacy of our civil justice system {[}..{]} and the
representativeness of juries is not merely an aspiration but a guarantee
under state and federal constitutions and statutes'' \cite[p.1]{hans2021}.
In particular, it has been shown that juries that reflect the full range
of community perspectives are in a position to incorporate these diverse
views into their fact finding. The best-known and best-documented examples concern the need to balance popular juries from the point of view of ethnic groups because it is believed that a jury composed mainly of members of the
same ethnic group tends to be favourable towards a defendant of the same
group, and unfavourable to a defendant from a different ethnic group.
For example, \citet{sommers2003} in a mock jury experiment, comparing the
deliberations of all-white and racially mixed juries, discovered that
diverse jury deliberations were more accurate, more expansive, and
longer. It was not simply that the minority jurors contributed new and
different information, the white jurors acted differently in all-white
versus mixed-race juries: they made fewer factual mistakes, and raised
more issues and evidence, during the deliberation. Moreover,
representative juries are more likely to be seen as legitimate decision
makers, which in turn contributes to public confidence in the justice
system. For all these reasons, ``courts should ensure that jury selection
procedures serve the goal of maximizing the representativeness of jury
pools and civil juries'' \cite[p.8]{hans2021}.

Similar considerations can be found in documents related to the design of research assessments. For instance, in the document regarding the recruitment of panels for the British RAE, it is recommended a fair representation of ``all affected parties'' involved in the decision-making process. It was stated that the selection of panellist has to ensure that ``the overall body of members reflects the diversity of the research community, including in terms of age, gender, ethnic origin, scope and focus of their home institution, and geographical location which represents the international reference on the subject'' \citep{ref2010}. The European Peer Review Guide of the European Science Foundation suggests that ``the goal should be to ensure availability of diverse viewpoints, scientific perspectives and scholarly thinking'' and that the criteria to be adopted for the selection of experts must also be the `diversity' that is expressed in terms of ``gender balance, scholarly thinking, background, geography,
turnover'' \citep{esf2011}. It appears that, as for legal juries, a panel composed by a cross-section of members of a scientific community may reach not only procedurally fair decisions, but also substantively better decisions \cite[p. 97]{elster}. From an empirical point of view, it is easy to test the fairness of a panel's composition in terms of observable characteristics of its members, such as gender, age or affiliation. It is instead much more complicated to test the fairness of its composition in terms of scholarly thinking, background, or other relevant features. In what follows, a test is proposed for fairness of a panel's composition based on the observation of intellectual and social connections among its members.

\section{The Italian research assessment exercises and the selection of panellists.}

In the Italian case, the question of the fairness of panel composition was mainly considered in terms of easily observable characteristics of panellists. The National Agency for the Evaluation of the University and Research (ANVUR) was charged in 2011 by a Ministerial Decree (DM) of realizing a newly designed research assessment, called VQR 2004-2010 (DM 17 of 15 July 2011). Afterwards, two other research assessments were designed and realized: the VQR 2011-2014 (DM of 27 June 2015) and the VQR 2015-2019 (DM of 25 September 2020). The three VQRs were mandatory and organized around research areas as defined by the National University Council. The assessment of each area was managed  and realized by a panel of scholars, the so called GEV (Group of Evaluation Experts). By and large, the areas were classified as ``bibliometric'' and ``non-bibliometric''. In both VQR 2004-2010 and VQR 2011-2014, the evaluation of bibliometric areas was conducted by a prevalent use of bibliometric algorithms, while for non-bibliometric areas it was realized by expert review. The VQR 2015-2019 adopted instead expert review, informed by bibliometrics, for all the research areas. The role of the panel was crucial in the three VQRs, since they defined for each area the specific rules for the evaluation. In particular panels decided bibliometric criteria, they defined the procedures for deciding which works should be evaluated with bibliometrics and which with peer review, they chose and coordinated the reviewers, they summarized the review reports, they evaluated in many cases directly the works submitted for evaluation \citep{baccini2016,Baccini_cjils,Baccini_De_Nicolao_2016}.

Given their crucial role, the question of panel composition is central in the quality and credibility of the research assessment. The above cited ministerial decrees defined also the procedures for the selection of panel members.  These procedures are described in detail in the Supplementary materials \ref{subsectionA1}. Here, it is sufficient to mention that the members of the committees for VQR 2004-2010 and VQR 2011-2014 were appointed directly by the ANVUR governing board mainly but not exclusively from among the Italian and foreign scholars who had applied in response to public calls to serve on the committees. Instead, the members of the panels for VQR 2015-2019 were selected exclusively by lot among those who applied as panellists. This institutional discontinuity introduced for the VQR 2015-2019 dramatically weakened the power of ANVUR governing board in the appointment of panels. It was the result of a political choice made by a government supported by a different majority than the previous ones. 

Indeed, the VQR 2004-2010 procedure was designed and completed by minister Maria Stella Gelmini of a center-right government led by prime minister Silvio Berlusconi, and by Francesco Profumo, minister of the technical government of Mario Monti. The VQR 2011-2014 was designed and completed by Minister Stefania Giannini of a center-left government led by prime minister Matteo Renzi. The VQR 2015-2019 was originally designed by minister Lorenzo Fioramonti, of the Five-stars Movement (an `anti-establishment' party) and center-left government led by prime minister Giuseppe Conte. Indeed, it can be conjectured that minister Fioramonti, in some occasions explicitly critical of previous research assessments, changed the rules for the appointment of panel members by following the public discussion developed during the previous years, which will be briefly illustrated below.

As anticipated, the literature on the procedural fairness suggests to compose panels in such a way that there is a fair representation of all affected parties involved in the decision-making process \citep{Leventhal1980}. Actually, the rules of the three VQRs defined generic criteria or thresholds for panel composition in terms of the observable characteristics of scholars: gender and affiliation, and research fields. As for gender, the first two VQRs requires a fair gender distribution, and VQR 2010-2014 fixed a threshold of about 33.3\% for women in the panel. As for affiliation, VQR 2004-2010 formally required a presence of 20\% of foreign scholars; VQR 2010-2014 a ``significant percentage'' without any explicit threshold; VQR 2015-2019 required 5\%. For Italian scholars, a not better specified fair distribution of panellists among universities and institutions is generally required in all the three VQRs. In the final reports of the assessments, ANVUR claimed that gender and affiliation criteria were met, presenting some data on the whole set of panellists, but no evidence was provided for each area panel \citep{ANVUR_2013,ANVUR_RF_2016,ANVUR_2022}. Hence, it may be that in some panels the criteria of fair composition in terms of gender and affiliations were not respected.

As for research fields, all three VQRs required the coverage of the research fields inside each Area, without any specific indication or thresholds. For the VQR 2004-2010, it was required that panels ``cover all the cultural and research lines within the areas'' \citep{ANVUR_2011}; in the VQR 2010-2014 and VQR 2015-2019 it was requested the ``coverage of the scientific-disciplinary sectors (SSD)'' \citep{ANVUR2015}. No specific indication or thresholds were defined. The final reports of VQR 2004-2010 and VQR 2010-2014 did not present data about the respect of the coverage of the research fields inside each area \citep{ANVUR_2011, ANVUR2015}; for the VQR 2015-2019 uncommented raw data about panel composition in terms of scientific-disciplinary sectors were published (https://www.anvur.it/attivita/vqr/vqr-2015-2019/gev/).

Moreover, the respect of fairness criteria about observable characteristics of panellists such as gender, affiliations and research field did not guarantee that panels had a fair composition in terms of diverse viewpoints, scientific perspectives and scholarly thinking. 

Indeed, the fairness of the composition of panels for the VQR 2004-2010 was questioned from its inception, by highlighting the lack of transparency in the members appointing procedures \citep{BACCINI2011}. The attention was especially focused on the panel for economics, statistics and business. This panel was composed not only by a small minority of women (16.7\%), but its appointed members were closely linked to each other by co-authorship relationships. In the economics sub-panel,  9 panellists out of 20 (45\%) were among the founders of a ultra-liberal party (``Fare per fermare il declino'') that participated in the political elections of 2013 by obtaining the 0.9\% of the votes and no representatives in parliament \citep{Baccini_2018}. The same problem was also documented for the VQR 2011-2014 during a conference organised at the Italian Parliament: the panel of economics was again largely composed by scholars of the same ultra-liberal party \citep{BACCINI_2016}. 
In the public debate it was noted that the absence of fairness in the composition of the panel for economics, statistics and business had probably minimized the voices of dissent with respect to the evaluation methods and rules adopted by the panel \citep{BACCINI2011, re2019}. It was noted also that there was a relevant precedent to consider: in the first experimental research assessment for the years 2001-2003, managed by CIVR (Steering Committee for Research Evaluation), Luigi Pasinetti, one of the panel member for economics, wrote a note of dissent documenting the absence of pluralism in evaluation \citep{Pasinetti}.  

\section{Research design}

The main aim of the paper is to analyze whether the selection procedure of the panellist for the Italian research assessments gave rise to a composition that fairly represents the intellectual diversity of the research community of economics, statistics and business. The analysis of the composition of panels in terms of gender or affiliations is important, but it is not sufficient. Indeed, a fair composition in terms of gender and affiliations does not guarantee that it is also respected a fair intellectual composition of the panels in terms of diverse viewpoints, scientific perspectives and scholarly thinking heterogeneity.

Unfortunately, the analysis of the intellectual composition of a panel is a very complex task for at least two reasons. First of all, it is difficult to classify panel members according to their intellectual perspective. Even if this classification is finally accomplished, it is difficult to evaluate if the panel composition reflects the intellectual diversity of the scholarly community at large. Instead of adopting a qualitative analysis of intellectual composition of panels and of scholarly communities of economists and statistics, this paper adopts a network analysis perspective. The basic idea is to explore   the intellectual and social compositions of the panels by observing connections between members. 

The interesting connections between scholars regard theoretical approaches, personal knowledge and economic-political visions. They are observed by building three different networks: a co-authorship network, a journal based network and an ``affinity network''.

A \emph{co-authorship network} permits to observe collaborations among scholars: in it two nodes representing two scholars are linked by a weighted edge if they have authored at least a paper together; the weight of the edge is proportional to the number of co-authored papers. More precisely, we adopted a so called \textit{ego}-coauthorship network approach (for a review see \cite{arnaboldi}). We started from the lists of panellists; the co-authorship network of a given panellist is defined as the weighted network formed by her/him and all her/his co-authors; each edge is weighted with a measure of the strength of the collaboration, usually, again, the number of co-authored paper. This approach is adopted when one is interested, as we are, in studying direct collaborations between a set of scholars, since it allows one to identify only (i) direct collaborations between members of the starting set of authors, in our case panels, and (ii) ties generated by two members of the starting set collaborating with a common co-authors.

Collaboration in research is a complex social phenomenon that has been systematically studied since the 1960s and co-authorship is the most tangible and well documented forms of scientific collaboration (for a review see \cite{kumar2015}). Co-authorship is considered a reliable proxy of research collaborations because co-authors cannot write a paper together unless some degrees of personal acquaintance exists between them.

However, there are many scholars who know one another or who are intellectually similar to some degree but have never collaborated by writing a paper. For this reason, to detect hidden connections, it is possible to look also for similar specialization, similar training and similar affiliations. 
A proxy of intellectual affinity among scholars can be obtained by observing the network of journals where scholar published their papers. 

In the \emph{network based on journals} two scholars (nodes) are considered connected if they have published in the same journal (edge). The relation between social and intellectual community gathered around economics and statistical journals is largely documented \citep{baccini_gingras,baccini_et_Al_2022}. Moreover, the choice of journals is so much important for career and promotion, especially in economics, that the observation of the journal portfolio of a scholar is often the only task done in recruitment processes and evaluation exercises \citep{heckman}. Scholars publishing in a same journal are sending the same signal to their peers about their achievements and positioning. The overlapping of journal portfolios of scholars can be considered as indicative that they are working in similar intellectual environments, that there is some sort of intellectual or theoretical similarity between them.  

Finally, the `\emph{affinity network}' is here defined as a generalized  affiliation network where two scholars (nodes) are connected if they are affiliated in the same university or in the same research centre, or if they studied at the same institution, or if they contributed to the same newspaper or blog (edge). This network is based on the hypothesis that, even if scholars have not published together or in the same journals, they can have a common set of relational patterns which can reflect theoretical or political similarities: not all economics departments and research centres carry out the same theoretical vision, and analogously magazines, newspapers and blogs have different editorial lines. 

These three networks will be analyzed for exploring the connections between the panellists, by considering appropriate quantitative indicators. In every case, in absence of a reasonable benchmark for the considered indicators, it is difficult to conjecture about the strength of the connections between members that corresponds to an unfair composition of the panel. 

In the absence of such a benchmark, a suitable control group representative of the research community of economics, statistics and business could be used for evaluating the fairness of the composition of panels. Theoretically, the best strategy would have been to build a control group for each of the three panels. By comparing the existing connections among members of each panel with those in the corresponding control group, it would be possible to assess whether or not the heterogeneity present in the research community was fairly represented in the panel. Unfortunately, the building of such control groups is very difficult, both theoretically and practically.

Indeed, as saw above, the composition of panels was the result of different two-step procedures. For the three panels the first step consisted in the definition of a list of eligible candidates through a worldwide public call to serve on the panels. As a consequence, the eligible scholars were self-selected from the world population of scholars. Scholars demanding for being part of panels were scholars who self-evaluate themselves as “prominent scholars”, and who accepted to work for a governmental agency in an administrative process. 
Moreover, foreign scholars were probably solicited to participate to the call by ANVUR or by other Italian scholars: as we documented in the Supplementary materials \ref{subsectionA1} the big majority of scholars with non-Italian affiliation have Italian first and last name. 

The second step of the procedure of appointment was instead different for the three VQRs. As anticipated, ANVUR governing board directly selected panel members for VQR 2004-2010 and VQR 2011-2014; in both cases ANVUR governing board appointed members by choosing them from the list of candidates, but also off the list in a number of cases that have not been disclosed. For the VQR 2015-2019 members were selected by lottery exclusively among the list of candidates.

Hence we can affirm that in VQR 2004-2010 and VQR 2011-2014 the final composition of panels is due in part to self-selection of scholars and in part to ANVUR choices, while in 2015-2019 the final composition is due only to self-selection.

From a practical point of view, the building of control groups for each of the three panels would require (i) random selection from all scholars active worldwide in economics, statistics and business in the year of panel appointment; (ii) the respect of generic criteria -- as described in Supplementary materials \ref{subsectionA1} --for the panel composition in terms of scientific production, gender, affiliations and geography. This strategy was practically unfeasible, given the huge amount of necessary information. If feasible, the adoption of this kind of control groups would give raise to different results for the three VQRs. Indeed, for VQR 2015-2019 an unfair composition would be due to self-selection mechanism. For VQR 2004-2010 and VQR 2011-2014 an unfair composition would be due to self-selection mechanism or to ANVUR choices or a combination of the two.

So, the strategy adopted in this paper consists in considering the panel for the VQR 2015-2019 as a sort of control group for the two preceding VQRs. 

We started by hypothesizing that there are no reasons to think that self-selection of candidates in the three VQRs happened in different ways. A rather rigourous test of this hypotesis could have consisted in building random panels from those who had applied for being member of the panels VQR 2004-2010 and VQR 2011-2014. A comparison of these two random panels with the one of 2015-2019 would have permitted to verify if self-selection operated differently in the three VQRs. Unluckily, the lists of candidates VQR 2004-2010 and VQR 2011-2014 were not publicly available.

If self-selection of candidates in the three VQRs happened in similar ways, self-selection did not generated \textit{per se} a different degree of unfairness in the composition of the three panels. Since in VQR 2015-2019 ANVUR governing board had no role in selecting panel members, a comparison of the composition of the panels of VQR 2004-2010 and VQR 2011-2014 with that of VQR 2015-2019 should allow us to isolate the effect of ANVUR governing board choices in the appointment procedures.

More explicitly, the panels for the VQR 2004-2010 and VQR 2011-2014 are considered as ``treatment groups'' and they are compared with the control group represented by the panel for the VQR 2015-2019. Differences in the structural properties of the three networks (co-authorship, journal based and affinity) between treatment groups and the control group are considered as indicative of unknown systematic biases in the choice of panellists introduced by ANVUR governing board choices. These biases might result in a more unfair composition of the two ``treated panels'' that the third.

In any case, the three groups are selected in different periods of time and the results of previous research assessments had probably an effect in the last one. For example, a bad evaluation of certain schools of thought could have led to a reduction of their funds and of scholars of these schools, as documented by \cite{lee2013} for the United Kingdom. Thus, it cannot be excluded that the level of heterogeneity in economics, statistics and business communities in 2011, when the panellists of VQR 2004-2010 were selected, was higher than the one in 2020, when the selection for the VQR 2015-2019 happened. This consideration would reinforce our research strategy, since the networks for the control group should be \textit{naturally} more concentrated than the ones for the VQR 2004-2010 and VQR 2011-2014.

The network analysis was conducted by using Pajek version 5.14 \citep{denooy2018}; visualizations are realized by VOSviewer version 1.6.15 \citep{van2020}. Data for replicating results and supplementary tables and figures (here after SM) are available here: 10.5281/zenodo.7244943.

\section{The co-authorship networks}

Three \textit{ego} co-authorship networks are built by considering the set of publications of all the members of each panel. In each network nodes are panel members and scholars who co-authored at least a paper with a panel member. Ties between nodes indicate direct collaboration between a pair of panellists, or collaboration of panel members with external scholars. Each network therefore includes direct co-authorships between pairs of panellists and indirect collaborations realized by writing a paper with a common co-author. Ties are weighted with the number of co-authored papers. The comparison of the structures of the three networks may permit to conjecture about the presence of diverse viewpoints, scientific perspectives and finally of scholarly thinking heterogeneity.

For observing collaborations among members before they met on the panel, their publications were retrieved from Scopus for the 25 years preceding the starting date of the research assessment exercises.  

The first observation is about the degree of overlapping of the three panels. 5 members of the panel 2004-2010 were appointed also as members of the panel 2011-2014 (Bartolucci F., Bertocchi G., Gambardella A., Ronchetti E., Schivardi F.). In particular the president of the panel 2011-2014 was also one of the member of the panel 2004-2010. In the co-authorship network 2011-2014 there are 10 other scholars that were also members of the panel 2004-2010 (Canova, Dardanoni, Dosi, Ellul, Frittelli, Jappelli, Peracchi, Rossi, Weber, Zamagni). In the panel 2015-2019 only one member was also a panellist in the panel 2011-2014 and one of the nodes of the co-authorship network of the panel 2004-2010 (Pagano). 

Table \ref{table:7} compares some structural statistics for the three coauthorship networks.

The networks are only slightly dissimilar in terms of number of nodes (authors) and links (coauthorships). The total number of co-authors and the average number of co-authors per panel member is growing from the first to the third panel. The greatest set of authors of the third panel produced only a lightly smaller number of papers than the other two panels. In the other two panels, especially the first, a smaller set of authors produced more papers: hence the degree of overlapping co-authorships is greater in the first two panels than in the third. 

\begin{table}[h!]
\caption{Basic statistics of the co-authorship networks.}
\label{table:7}
\begin{tabular}{|l|c|c|c|}
\hline
& \textbf{Panel 2004-2010} & \textbf{Panel 2011-2014} & \textbf{Panel
2015-2019} \\
\hline
N. of panel members & 36 & 31 & 40 (37)$^*$\\
Publication years & 1987-2011 & 1991-2015 & 1996-2020 \\
N. of co-authors of panel members& 781 & 922 & 1,271 \\
N. of co-authors per panel member & 21.7 & 29.7 & 34.0 \\
N. of papers & 1,190 & 1,188 & 1,079 \\
N. of papers per panel member & 33.1 & 38.3 & 29.2\\
Number of edges between authors & 1,801 & 2,829 & 5,678 \\
Minimum weight of an edge & 1 & 1 & 1 \\
Maximum weight of an edge & 47 & 142 & 31 \\
Number of edges with weight equal to 1 & 1,315 & 2,178 & 4,304 \\
Number of edges with weight equal to 2 & 251 & 337 & 880 \\
Number of edges with value greater than 2 & 235 & 314 & 494 \\
Density & 0.005 & 0.006 & 0.007 \\
Average Degree & 4.6 & 6.1 & 8.9 \\
Number of components & 12 & 17 & 25 \\
Percentage of realized components over maximum & 33.3 & 54.8 & 62.5 \\
\hline
\end{tabular}
\begin{quotation}
\footnotesize
$^*$ Three panellists were dropped from the analysis because no information was retrieved from Scopus (Cori Enrico, De Vincentiis Paola and Pisoni Pietro Maria).
\end{quotation}
\end{table}

In a co-authorship network, the degree of a node representing a scholar indicates the number of its co-authors. 
The degree distributions of the three networks (available in the Supplementary Materials) are statistically different, according to Kolmogorov-Smirnov test (first panel-second panel:  \(D= 0.125 (p=3.48e-06)\); first panel-third panel: \(D = 0.31 (p=0.00e+00)\); second panel-third panel: \(D= 0.213 (p=0.000e+00)\); $p$-value adjusted for multiple comparison). In particular  
the coauthorship network for the third panel appears as radically different: it has the lowest share of scholars with only 1 or 2 co-authorships (32$\%$ against 41$\%$ of the panel 2004-2010 and 54$\%$ of the panel 2011-2014) and the biggest share of scholars with more than 22 co-authors (11.64\% against 1.41\% of the panel 2004-2010 and 4.37\% of the panel 2011-2014). 

The most central scholars can be identified by computing their betweenness centrality \citep{wasserman_faust}. The distributions of betwenness centralities in the three panels are not statistically different, but their comparison gives some qualitative indications. The maximum value of betweenness is $0.25$ for the first panel, $0.099$ for the second and $0.045$ for the third panel. The share of nodes with zero betweenness centrality, i.e. nodes that are coauthors of only one member of the panel, is similar in the three networks, but the first and the second networks have higher share of nodes with relatively high value of betweenness. In the third panel only 16 scholars (1.26\%) have a betweenness higher than $0.002$, against 50 scholars (6.40\%) in the first panel and 31 scholars (3.36\%) in the second panel. The list of these most central scholars are reported in Table \ref{table:A1}, \ref{table:A2} and \ref{table:A3}.
Again, 10 scholars appears in both the lists of the most central scholars of the first two panels (Bartolucci, Bertocchi, Dosi, Gambardella, Guiso, Jappelli, Lippi, Pagano, Peracchi, Schivardi). Only one of these scholars is also among the most important scholars of the panel 2015-2019 (Pagano).

\begin{sidewaysfigure}
\centering
\vline
\subfloat [Panel 2004-2010: 12 components]{\includegraphics[scale=0.5]{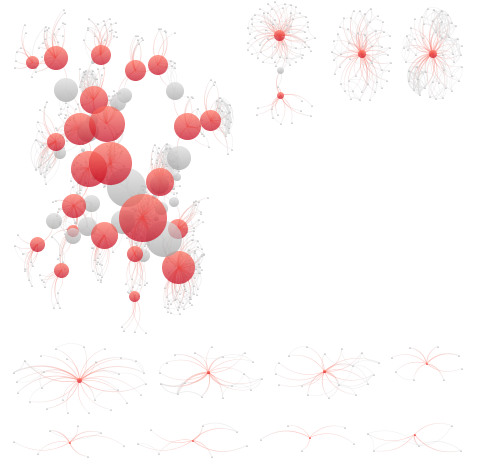}}
\vline
\subfloat [Panel 2011-2014: 17 components]{\includegraphics[scale=0.5]{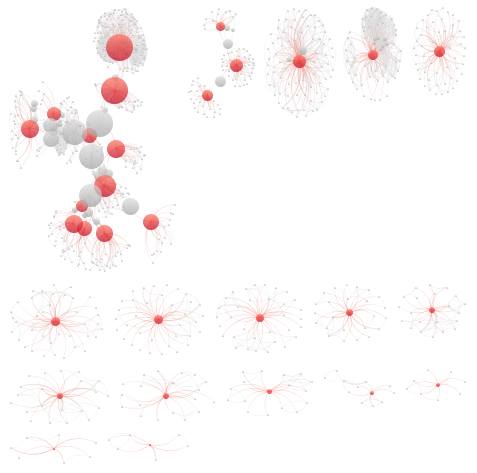}}
\vline
\subfloat [Panel 2015-2019: 25 components]{\includegraphics[scale=0.5]{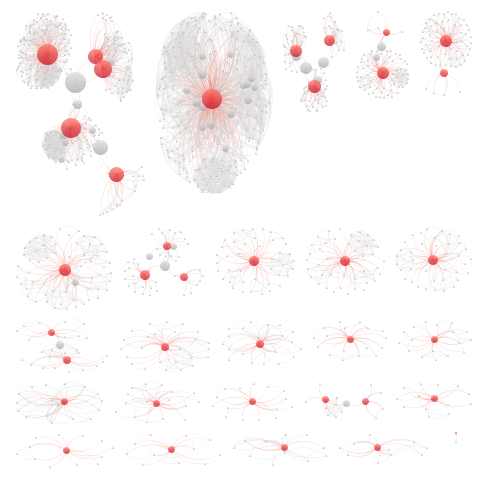}}
\vline
\caption{Components in co-authorship networks. Size of vertices is proportional to betweenness centrality and red nodes are panel members. An edge between two nodes indicates that the two scholars wrote together at least one paper. Thickness of edges is proportional to the number of common papers.}
\label{figure:4}
\end{sidewaysfigure}

Figure \ref{figure:4} shows that each of the three networks is partitioned in components, i.e. separated sub-networks \citep{newman}. It is straightforward to interpret the components of each network as communities of scholars with relatively strong co-authorship relations. Please note that the size of networks and the fact that they are naturally partitioned in components, makes it unnecessary to resort to community detection algorithms \citep{newman}.  

In a co-authorship network, the maximum level of fragmentation is reached when the number of components is equal to the number \(n\) of panellists. In this configuration each component of the network includes a member of the panel and all her/his co-authors; scholars belonging to different components are not co-authors. The ratio between the actual numbers of components of a network and the maximum can be considered as an indicator of the fragmentation of the panel, with values in the range \([1/n,1]\). 

If the co-authorship networks of the panel 2004-2010 and of the panel 2011-2014 are less fragmented, i.e. they have less and larger components than the panel 2015-2019, this can be considered as a clue of their unfair composition.

The penultimate row of Table \ref{table:7} indicates that the first panel has the lowest number of components (12), while the third panel has the highest (25). The last row of Table \ref{table:7} reports that the panel 2015-2019 is the most fragmented with a ratio of actual components over the maximum of 0.625 against 0.333 and 0.548 of, respectively, the first and the second panel. 

Fig.\ref{figure:4}a shows that the panel 2004-2010 is characterized by a big component containing 512 nodes, i.e. more than 65\% of the nodes of the network, and 24 panel members out of 36 (66\%). It contains also 10 other components each gathering one panel member and the set of her/his coauthors; only one component contains two panel members and their coauthors. Fig.\ref{figure:2} of the Appendix represents the biggest component, by showing clearly the central role that some scholars, who are not panel members, play in the construction of the network. 

As anticipated, the 2011-2014 co-authorship network is a bit more fragmented than the first one. Table ref{table:4} reports the frequency distribution of the components. Also in this case, as shown in Fig.\ref{figure:4}b, the network is characterized by only one big component with 403 nodes, i.e. more than 43\% of the scholars in the network. In this big component, there are 13 panel members representing the 42\% of the panellists. 17 components gather each one panel member and the set of her/his coauthors; only one component contains three panel members and their coauthors. Fig.\ref{figure:3} of the Appendix draws the biggest component illustrating, also in this case, the central role of non-panellists in the structure of the network. 

The 2015-2019 co-authorship network depicted in Fig.\ref{figure:4}c is the most fragmented. In this case, the largest big components contains only 324 nodes, i.e. the 25\% of the network, and only 5 panellists out of 40 (13\%). In the network there is another big component that contains 260 nodes (20,4\%), but only 1 panel member (Stingo F.), who has an exceptional number of co-authorship connections (259). There are also 6 components gathered around 2 or 3 panel members, and with variable size. 

More detailed data about the three co-authorship networks and their components are available in the Supplementary Materials.

In sum, the comparison of the three co-authorship networks shows that the first two panels are structurally different from the third. The first one has few central scholars mutually connected in one single large cluster that contains more than half of the pannelists and of their co-authors. The second network is completely similar to the first one, and shares with it a same group of central nodes. In contrast, the network for the 2015-2019 panel is fragmented into smaller components, consisting mainly of one to three panellists and their co-authors; no scholars show particularly high centrality values, and very few are in common with other networks. 

\section{The journal based networks}

The analysis of the journal based networks aims to investigate whether the panel members have published in the same or in different journals. If two authors have published articles in the same set of journals, then they are in some sense similar in terms of topics or theoretical approaches or methodologies. Also in this case the analysis consists in comparing the networks of the panels 2004-2010 and 2011-2014 with the control group represented by the panel 2015-2019. If the sets of journals of the first two panels are narrower than the set of the control group, this may be considered a clue indicating that ANVUR's selection introduced unfairness into the panel composition.

The dataset for journal based networks is the same used for the co-authorship networks. For each panel, the starting point is a bipartite network where panel members are linked to the journals where they have published at least a paper. For each panel, the one mode projection is then derived, where nodes are the panel members and the edge between two panel members is weighted according to the number of journals where they both published at least a paper. 

Table \ref{table:15a} reports the basic statistics of the three bipartite networks. Despite the three networks are similar in terms of panel members, the sets of journal is growing from the first panel (360 journals), to the second (467), to the third (566). In particular the first panel has the lowest number of journals per panellist. The members of the third panel on average published their work in a greater sets of publishing outlets than the first two.

\begin{table}[h!]
\centering
\caption{Basic statistics of the journal based networks}
\label{table:15a}
\begin{tabular}{|l|c|c|c|}
\hline
& \textbf{Panel 2004-2010} & \textbf{Panel 2011-2014} & \textbf{Panel
2015-2019} \\
\hline
N. of journals & 360 & 467 & 566 \\
N. of panel members & 36 & 31 & 37 \\
Number of edges & 511 & 675 & 721 \\
Number of journals per panel member & 10 & 15.1 & 15.3 \\
Density {[}2-Mode{]} & 0.040 & 0.046 & 0.034 \\
\hline
\end{tabular}%
\end{table}

As a consequence, in the one-mode projection networks of scholars density and average degree tend to be lower in the third panel than in the first two, as it is reported in Table \ref{table:15}.

\begin{table}[h!]
\caption{Journal-based networks: basic statistics of the one-mode projection networks of scholars.}
\label{table:15}
\centering
\begin{tabular}{|l|c|c|c|}
\hline
& \textbf{Panel 2004-2010} & \textbf{Panel 2011-2014} & \textbf{Panel
2015-2019} \\
\hline
N. of panel members & 36 & 31 & 37 \\
Number of edges & 148 & 137 & 137 \\
Lowest weight of an edge & 1 & 1 & 1 \\
Highest weight of an edge & 8 & 13 & 14 \\
Number of edges with weight equal to 1 & 86 & 59 & 86 \\
Number of edges with weight equal to 2 & 32 & 18 & 35 \\
Number of edges with weight greater than 2 & 30 & 60 & 16 \\
Density & 0.228 & 0.285 & 0.200 \\
Average Degree & 8.2 & 8.8 & 7.4 \\
Islands & 3 & 2 & 3 \\
Number of off-island panellists & 9 & 11 & 22 \\
Percentage of off-island panellists & 25.0\% & 35.5\% & 59,5\% \\

\hline
\end{tabular}%
\end{table}

In these network the degree of a panel member is the number of other panellists who published at least a paper in at least a journal where he or she also published. In the first panel, the president of the panel T. Jappelli and A. Bisin have the maximum degree, being linked to other 17 panellists; the president of the second panel G. Bertocchi has a degree of 17, the second highest degree after L. Sarno; in the third panel M. Piva has the highest degree of 22.

The search for clusters of panellists in the projection networks is conducted by using a simple edge-cut technique based on edge weights. The algorithm detects ``islands'' i.e. maximal subnetworks of nodes connected by edges with values greater than the ones of the edges linking nodes outside the subnetwork \citep{denooy2018}. For interpreting the results, consider that in a panel with the maximum level of diversity, none of the members belongs to an island, i.e. panel members tend to publish their articles in different journals and the publication of articles in a same journal is not systematic or a rare event. A simple indicator of intellectual diversity of the panel is therefore the share of panel members who are not clustered in an island; this share tends to 100\% for a panel with a maximum diversity.   

The search for islands (of minimum size 1 and maximum size of 3/5 of the number of panellists, respectively 21, 18 and 22) in the three networks individuated a similar numbers of clusters (3, 2 and 3), but with different configuration.  As reported in Tab. \ref{table:15}, in the third panel about 60\% of members are not part of any island, against about 42\% in the first panel and 45\% in the second. The main difference emerge when the largest island of each network is analyzed, as reported in Table \ref{table:16}. In the third panel the largest island gathers only 6 panellists (16.2\%) while in the first and second panel the biggest islands gather respectively 21 members, representing the 58.3\% of the panel, and 17 members for a 54.8\%. Moreover, the largest island of the first panel is less centralized with respect to the other, by showing that in this first panel all the members have similar roles in structuring the island, while, especially in the third, there is a prevailing role of a very small set of nodes. In Tables \ref{table:10}, \ref{table:12} and \ref{table:14} are reported the name of the scholars, the belonging island and the betweenness values with the connected ranking for each panel. 

\begin{table}[h!]
\caption{Basic statistics of the largest island in the one-mode projection networks of scholars.}
\label{table:16}
\centering
\begin{tabular}{|l|c|c|c|}
\hline
& \textbf{Panel 2004-2010} & \textbf{Panel 2011-2014} & \textbf{Panel
2015-2019} \\
\hline
N. of panel members & 21 & 17 & 6 \\
Percentage of totale panel members & 58.3\% & 54.8\% & 16.2\%\\
Number of edges & 58 & 25 & 8 \\
Lowest weight of an edge & 2 & 5 & 3 \\
Highest weight of an edge & 7 & 10 & 14 \\
Number of edges with weight equal to 1 & 0 & 0 & 0 \\
Number of edges with weight equal to 2 & 30 & 0 & 0 \\
Number of edges with weight greater than 2 & 28 & 25 & 8 \\
Density1 {[}loops allowed{]} & 0.263 & 0.173 & 0.444 \\
Average Degree & 5.523 & 2.941 & 2.666 \\
\hline
\end{tabular}%
\end{table}

\begin{sidewaysfigure}
\centering
\vline
\subfloat [Panel 2004-2010: 3 islands]{\includegraphics[scale=0.5]{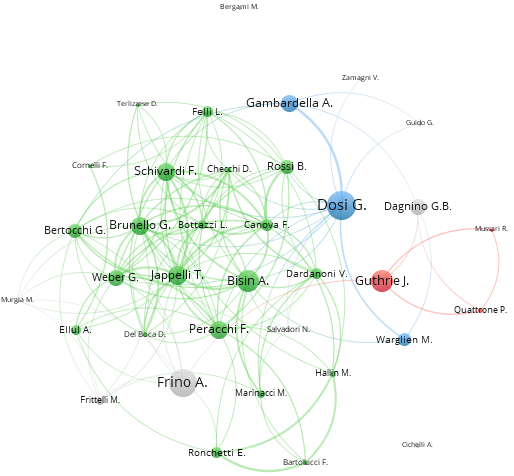}}
\vline
\subfloat [Panel 2011-2014: 2 islands]{\includegraphics[scale=0.5]{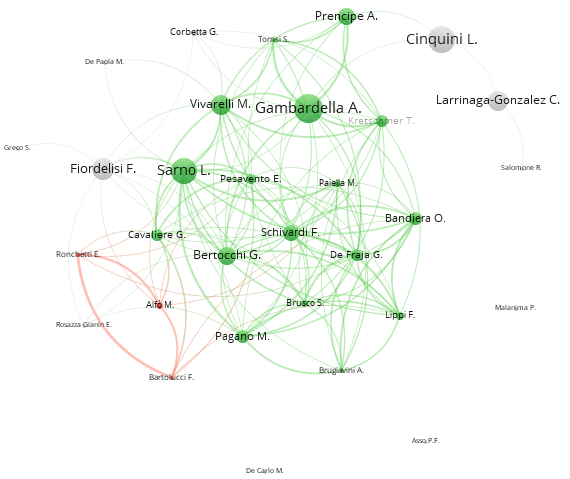}}
\vline
\subfloat [Panel 2015-2019: 3 islands]{\includegraphics[scale=0.5]{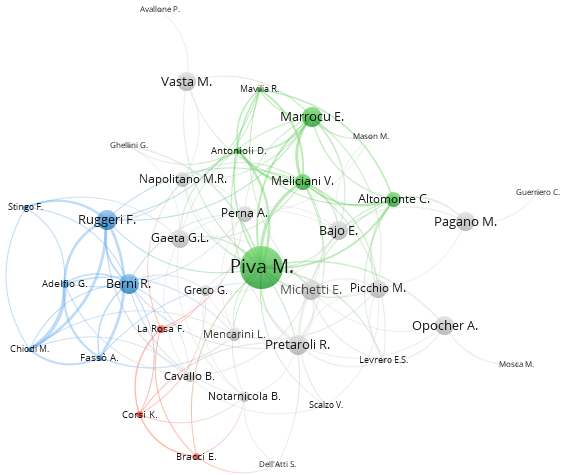}}
\vline
\caption{The clusters of scholars in the islands of the journal based networks. Each node is a panel member. Size of nodes is proportional to their betweenness centrality. An edge between two nodes indicates that the two members wrote papers in at least a same journal. Thickness of the links is proportional to the number of common journals. The color of nodes indicates different islands. Grey nodes are not part of any island.}
\label{figure:6}
\end{sidewaysfigure}

As visualized in Figure \ref{figure:6}(a), only 2 members of the 2004-2010 panel are isolated (Cicchelli and Bergami), i.e. they wrote in journals where none of the other panellists wrote. The biggest green island is composed by nodes sharing a relatively high betweenness centrality, by confirming that the big part of panellists contribute similarly to the network structure. Bisin has the highest betweenness centrality followed by the president of the panel Jappelli;  only three scholars of the green island have betweenness centrality lower than 0.002 (Cornelli, Del Boca and Terlizzese), by occupying a peripheral role in the network.

The blue island is composed by only three nodes with relatively high betweenness centrality. The red island also hosts only three scholar, but only one of them has a relatively high betweenness centrality. 
Among the scholars that are not part of an island, 6 have zero or almost zero betweenness centrality; Frino has the second largest value, i.e. he wrote in many different journals in which other panellists wrote, but the number of journals shared with others is not high enough to attract him to an island. 

The focus on the connection between panel members through journals enriches the information obtained by observing co-authorships, by showing different centrality ranking and different clusters. In particular, Ronchetti, Bisin, Bottazzi, Rossi, who were isolated in the co-authorship network, are now part of the green island gathering most of the panel. This indicates, for example, that Bisin, while not collaborating directly with the other group members, wrote in the same journals as most of the other participants.   
In contrast, the panellists of the red island are completely integrated in the biggest component of the co-authorship network. Dosi, for example, did not share many journals with other members of the panel, but it is linked by direct (co-authorship) or indirect (common co-authors) collaborations with most of the panel. 

In the 2011-2014 panel, reported in Figure \ref{figure:6}(b), 3 panellists appear as isolated (Asso, De Carlo, and Malanima). Also in this network, the biggest green island is structured by nodes with similar and relatively high betweenness centrality. The green island gathers 17 scholars, i.e. about 55\% of the members. Gambardella is the scholar with the highest betweenness centrality; the president of the panel (Bertocchi) has the fourth highest centrality. 
The red island gathers three panellists with relatively low betweenness centrality and relatively peripheral position in the network. Among the 11 scholars that are not part of an island, 7 have a zero or almost zero betwenness centrality, while Cinquini has the second highest value of the network. 
Also for this second journal based network, the clustering structure is only slightly different with respect to the co-authorship network. In particular, with the only exceptions of Alfò and Bartolucci, all the panellists belonging to biggest component of the co-authorship network are now part of the green island. It grouped also six panellists (Sarno, Vivarelli, Cavaliere, De Fraja, Kretschmer, Pesavento) who were not in the biggest component of the coauthorship network. Inversely, two of the three members of red island were part of the biggest component in the co-authorship network. 

The main features of the 2015-2019 panel visualized in Figure \ref{figure:6}(c) are the absence of a big island, and the presence of a majority of nodes (22 out of 37) that are no part of any islands. The analysis of the connections through journals shows few differences with respect to the co-authorship network. The green island includes six panellists who were grouped in two small components in the co-authorship network. The blue island gathers six scholars too: four were grouped in the biggest component of the co-authorship network and the other two were isolated (Berni and Stingo). Two of the three nodes of the red island (La Rosa and Corsi) were grouped together also in a component of the co-authorship network. 14 isolated scholars in the co-authorship network do not belong to any island. 

In sum, the analysis of the journal-based networks shows that the first two panels are different from the third one. In the first two panels, a majority of members published on a relatively small set of journals, and only a small minority of them appears to have no systematic connections with others members. In contrast, the vast majority of members in the third panel did not wrote systematically in the same journals. The third panel appears as characterized by greater diversity in terms of publishing outlets than the first two panels.

\section{The affinity networks}

The analysis of the affinity networks aims to investigate whether panel members have studied or are affiliated with the same research centres and universities, whether they have published non-specialized articles in the same newspapers, magazines or blogs. Common affiliations indicate then they probably have personal ties or theoretical or political affinities. A panel characterized by intellectual diversity has members that studied in different universities, with different affiliations, and writing in different magazines, newspapers and blogs. Also for the affinity networks, the analysis consists in comparing the networks of the three panels, by considering the third one as a control group.

The dataset is built by considering panel members and the most central nodes in each co-authorship network, i.e. the coauthors of panellists with a betweenness centrality value larger than 0.002. They are reported in Table \ref{table:A1}, Table \ref{table:A2}, and Table \ref{table:A3}. This choice permits to pay attention to the ties between members and their coauthors. The basic idea is that a person, connected to people who in turn are not directly connected, can mediate with each other and profit from mediation \citep{denooy2018}. In our case, people acting as bridges in connecting panellists in the co-authorship network, could be an additional element in reducing panel diversity: if these bridges come from the same universities or research centres or publish in the same newspapers, they could spread the same vision.

The \textit{curriculum vitae} of the panellists and of their more central co-authors were collected online from 17/02/2021 to 09/04/2021 and were manually processed to derive the following information: institutions where they graduated (maximum 2); institutions where they did MSc/MA and PhD (maximum 2); universities where they declared affiliations (maximum 2); declared affiliation to research centres (maximum 5); magazines, newspapers and blogs in which they wrote (maximum 5). Hereinafter all these entities are refereed to as ``affiliated institutions''. 

The affinity networks are bipartite networks where panel members and their more central co-authors are linked to affiliated institutions. From each of these bipartite networks is possible to build two projections. In the projection network of scholars, two scholars are connected if they have at least a common affiliation, and the weight of their link is proportional to the number of their common affiliations. In the projection network of institutions, two institution are linked if they both have an affiliation by at least one same scholar, and the weight of their link is proportional to the number of common scholars. 

Table \ref{table:23} reports basic descriptive statistics of the three bipartite affinity networks. The panel 2004-2010 has the highest number of scholars, since it includes the biggest number of central coauthors of panel members.
The basic statistics of the one-mode projection networks of scholars reported in Table \ref{table:24} and of affiliated institutions reported in Table \ref{table:25} reveal that the third panel is structurally different from the others. 

\begin{table}[h!]
\centering
\caption{Basic statistics of the affinity networks}
\label{table:23}
\begin{tabular}{|l|c|c|c|}
\hline
& \textbf{Panel 2004-2010} & \textbf{Panel 2011-2014} & \textbf{Panel
2015-2019} \\
\hline
N. of scholars & 58 & 40 & 44 \\
N. of coauthors of panellists & 22 & 9 & 4\\
N. of affiliated institutions & 191 & 147 & 171 \\
Number of edges & 426 & 306 & 282 \\
Number of scholars per institution & 0.30 & 0.27 & 0.25 \\
Number of institutions per scholar & 3.29 & 3.67 & 3.88 \\
Density {[}2-Mode{]} & 0.038 & 0.052 & 0.037 \\
\hline
\end{tabular}%
\end{table}

In particular for the projection networks of scholars (Figure \ref{table:24}), the third panel has lower average degree and density than the other. Moreover, the third panel has also systematically lower edge weights, i.e. the number of common affiliated institutions between pairs of scholars is systematically lower in the third panel than in the others. 

Analogously, for the projection networks of affiliated institutions (Figure \ref{table:25}), the third panel appears as less connected: the number of edges is lower and very few institutions are linked by more than two affiliated scholars. As a consequence, in the third panel the average degree and density are lower than in the other two panels. 

\begin{table}[h!]
\caption{Basic statistics of the projection network of scholars}
\label{table:24}
\centering
\begin{tabular}{|l|c|c|c|}
\hline
& \textbf{Panel 2004-2010} & \textbf{Panel 2011-2014} & \textbf{Panel
2015-2019} \\
\hline
Number of scholars & 58 & 40 & 44 \\
Number of edges & 551 & 345 & 185 \\
Lowest weight of edges & 1 & 1 & 1 \\
Highest weight of edges & 7 & 7 & 5 \\
Number of edges with weight equal to 1 & 272 & 172 & 152 \\
Number of edges with weight equal to 2 & 129 & 95 & 24 \\
Number of edges with weight greater than 2 & 150 & 78 & 9 \\
Average Degree & 19.000 & 17.250 & 8.409 \\
Density & 0.327 & 0.431 & 0.191 \\
Number of islands & 4 & 2 & 7 \\
\makecell[l]{Largest number (and \%) of important vertices \\ in the same island} & \makecell{15 \\ (88.2\%)}  & \makecell{14 \\ (82.3\%)}  & \makecell{4 \\ (23.5\%)}  \\
Number of off-island scholars & 36 & 23 & 22 \\
Number of off-island important scholars & 2 & 3 & 7 \\
\hline
\end{tabular}%
\end{table}

\begin{table}[h!]
\caption{Basic statistics of the network of affiliated institutions}
\label{table:25}
\centering
\begin{tabular}{|l|c|c|c|}
\hline
& \textbf{Panel 2004-2010} & \textbf{Panel 2011-2014} & \textbf{Panel
2015-2019} \\
\hline
Number of affiliated institution & 191 & 147 & 171 \\
Number of edges & 1277 & 944 & 889 \\
Lowest weight of edges & 1 & 1 & 1 \\
Highest weight of edges & 15 & 12 & 4 \\
Number of edges with weight equal to 1 & 1083 & 798 & 845 \\
Number of edges with weight equal to 2 & 124 & 97 & 37 \\
Number of edges with weight greater than 2 & 70 & 49 & 7 \\
Average Degree & 13.371 & 12.843 & 10.397 \\
Density & 0.070 & 0.087 & 0.061 \\
Number of islands & 8 & 2 & 8 \\
\makecell[l]{Largest number (and \%) of important vertices \\ in the same island} & \makecell{15 \\ (88.2\%)}  & \makecell{8 \\ (47.1\%)}  & \makecell{11 \\ (64.7\%)}  \\
Number of off-island institutions & 153 & 131 & 131 \\
Number of off-island important institutions & 2 & 9 & 0 \\
\hline
\end{tabular}%
\end{table}

A better understanding of the features  of the bipartite affinity networks can be achieved by adopting a two-step analysis. In a first step, the sets of important scholars and affiliated institutions are searched by an algorithm based on eigenvector centrality \citep[p. 159]{newman}: a scholar or an institution is important if it is linked to other important scholars or affiliated institutions in the network. The lists of the most important scholars and institutions in the three networks are reported in Tables \ref{table:17}, \ref{table:19}, \ref{table:21}. 
The second step consists in looking for clusters, separately, in the networks of scholars and in the networks of affiliated institutions through the island algorithm, by fixing the minimum size to 1 and the maximum size to 17. It is then possible to observe how many of the 17 most important scholars/institutions are part of a same island. In this settings, the maximum degree of diversity is reached when all the most important scholars/institutions belong to different islands.

The basic statistics of the scholar networks are in Table \ref{table:24}; the number and dimensions of the clusters are reported in Tables \ref{table:18}, \ref{table:20} and \ref{table:22}.
The third panel appears as very different from the others. 
In the third panel, a big part of scholars (50\%) and important scholars (41.2\%) are not part of any island; moreover, the other are dispersed in 7 islands with a maximum of 4 scholars clustered together in a same island. The other two panels are characterized instead for having not only a smaller number of islands, but mainly for having a big central island gathering the big part of the important scholars. Moreover, in the first two networks co-authors of panellists play a very central role in structuring the affinity network, while in the third they have mainly a peripheral position. 

\begin{sidewaysfigure}
\centering
\vline
\subfloat [Panel 2004-2010: 4 islands]{\includegraphics[scale=0.5]{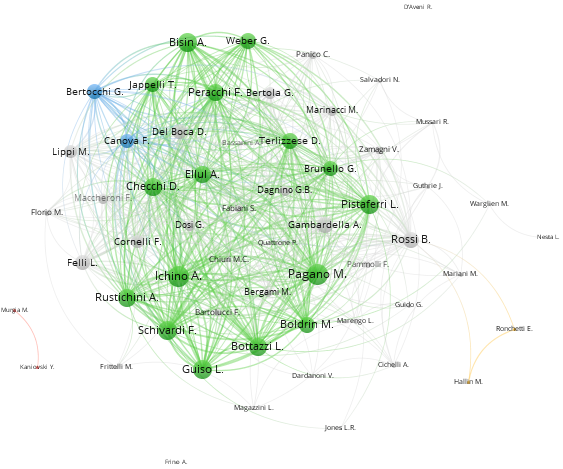}}
\vline
\subfloat [Panel 2011-2014: 2 islands]{\includegraphics[scale=0.5]{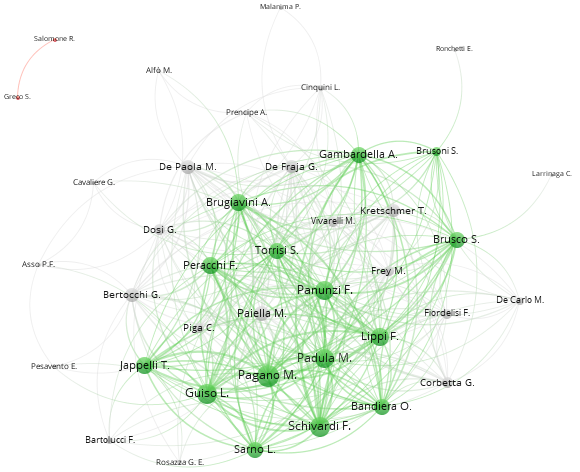}}
\vline
\subfloat [Panel 2015-2019: 7 islands]{\includegraphics[scale=0.5]{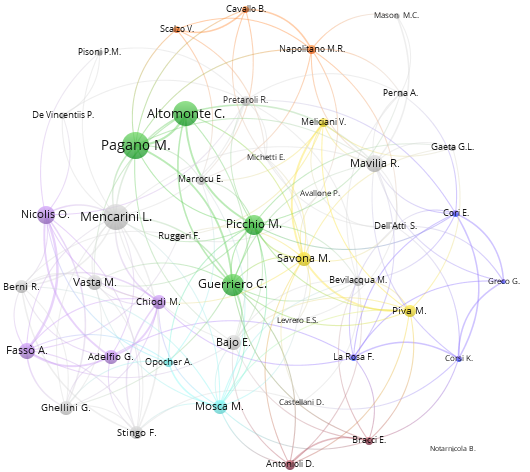}}
\vline
\caption{The clusters of scholars in the islands of the affinity networks. Each island has a different colour, and the dimension of vertices is proportional to eigenvector centrality in the affinity network. An edge between two nodes indicates that the two scholars have at least one common affiliation. Thickness of edges is proportional to the number of common affiliations.}
\label{figure:7}
\end{sidewaysfigure}

In particular, in the 2004-2010 affinity network, the network is composed by 58 scholars: 36 panel members and 22 co-authors. Among the 17 most important scholars 6 are coauthors of panel members; 4 of these coauthors are among the top five most important scholars. The biggest green island in Figure \ref{figure:7}(a) gathers 16 scholars, 15 of which are among the 17 most important scholars: near all the most important scholars share common affiliations.  

The 2011-2014 network of scholars, drawn in Figure \ref{figure:7}(b), is composed by 31 panel members and 9 co-authors. Among the 17 most important scholars, 5 are coauthors of panel members and 3 are in the top five. It contains  two islands. As in the first panel, the biggest island gathers 15 scholars, 14 of which important,  the most important scholars have a high degree of common affiliations.

Finally, the network of scholars for the third panel is composed by 40 panel members and 4 coauthors. It is represented in Figure \ref{figure:7}(c). Among the 17 most important scholars only 2 are co-authors of panel members and none of these is in the top five. 22 scholars (41\%) are not part of any island; 7 important scholars out of 17 are not part of any island. The remaining 22 scholars are dispersed in 7 islands of maximum size 4 scholars. Three of these islands have zero important scholars, three have only one important scholar and two island of size 4 are populated by important scholars only.  

As for the overlapping of the important scholars among the three affinity networks, from Tables \ref{table:17} \ref{table:19}, \ref{table:21} appears that 6 scholars (Gambardella, Guiso, Jappelli, Pagano, Peracchi, Schivardi) are both in 2004-2010 and in 2011-2014 networks, including the president of the first panel. Among these, Guiso is the only one who has never been a panel member. Only one panellist of the 2015-2019 panel (Pagano) appears also among the important vertices of scholars of the previous two panels. 

The projection network of affiliated institutions also appears structurally different for the third panel than in the others. The only common trait among the three is that Bocconi University in Milan is the university with the highest eigenvector centrality. But while in the first two panels Bocconi University is the center of the island that concentrates most of the most important connections and scholars, in the third panel Bocconi's island is small in size and flanked by other islands of not much different size even in terms of important vertices, as it is shown in Tables \ref{table:18} \ref{table:20}, \ref{table:22}. 
In particular, the 2004-2011 network of affiliated institutions drawn in Figure \ref{figure:8}(a) shows that 15 of the 17 most important institutions are clustered in the same island. As reported in Tables \ref{table:17}, the other important vertices of the largest island gathers American economic policy think tank (CEPR and NBER), universities research centres (EIEF, strictly linked to the Bank of Italy), other non-academic institutions (Bank of Italy), newspapers (IlSole24ore, Il Foglio) and blogs (lavoce.info).

\begin{sidewaysfigure}
\centering
\vline
\subfloat [Panel 2004-2010: 8 islands]{\includegraphics[scale=0.5]{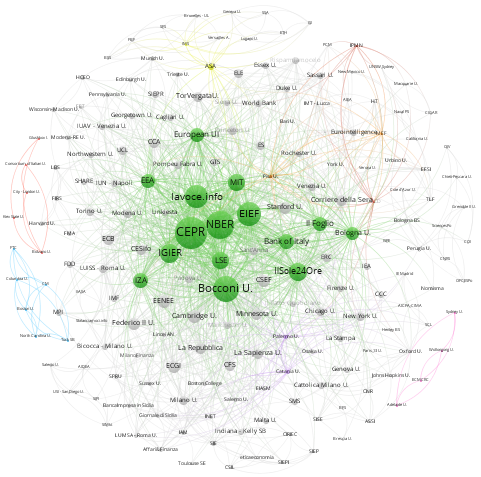}}
\vline
\subfloat [Panel 2011-2014: 2 islands]{\includegraphics[scale=0.5]{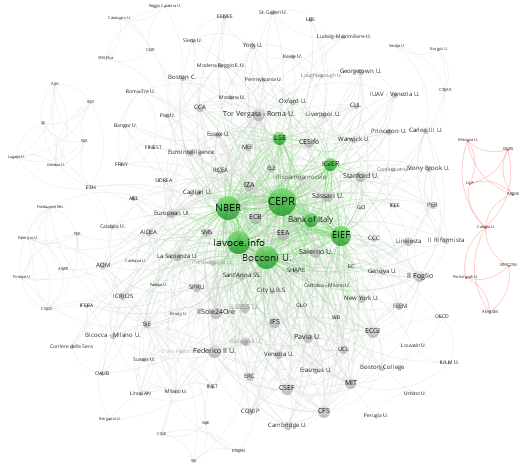}}
\vline
\subfloat [Panel 2015-2019: 8 islands]{\includegraphics[scale=0.5]{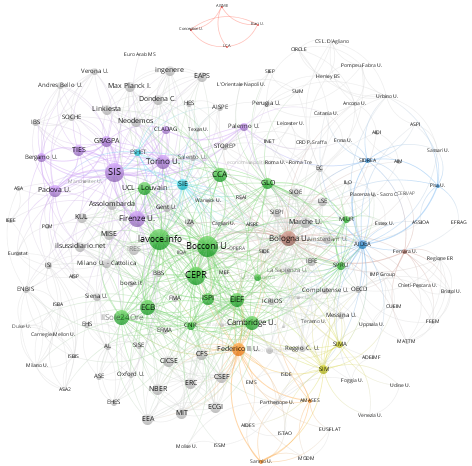}}
\vline
\caption{The clusters or islands of affiliated institutions for the three panels. Each island has a different colour, grey nodes do not belong to any island and the size of vertices is proportional to eigenvector centrality in the affinity network. An edge between two nodes indicates that the two affiliations are connected at least by one scholars. Thickness of edges is proportional to the number of common scholars.}
\label{figure:8}
\end{sidewaysfigure}

The 2011-2014 network of affiliated institutions, drawn in Figure \ref{figure:8}b, is very similar to the previous one. As reported in Table \ref{table:19}, the green island collects only important institutions. Essentially, the green island is a concentrated replica of the biggest island detected for the 2004-2010 panel, bringing together only one Italian university: Bocconi and its IGIER research centre, the Bank of Italy and the research center EIEF it founded, the blog lavoce.info and the American economic policy think tanks CEPR and NBER. The red island collects only not important vertices. 

The control group is drawn in Figure \ref{figure:8}(c). It is much more fragmented since the institutions form 8 islands: six of these islands, with a limited size of 2 to 4 nodes, have zero or only one important vertex. The biggest green island gathers 15 institutions, 11 of which are important institutions. The second biggest blue island contains 9 institutions, 4 of which are important. The list of important vertices of the biggest island (Table \ref{table:21}) is the same of the biggest islands of the two previous panel. However, only 4 important scholars are connected to this island. The second biggest island gathers three Italian universities and a scholarly society (the Italian statistical society).

In sum, if we consider the affinity network of the third panel as the benchmark for judging about the degree of diversity, the networks of the panel 2004-2010 and of the panel 2011-2014 appear as very far from the benchmark. Indeed, they are characterized by the prominent presence of a bulk of important scholars affiliated to a bulk of important affiliated institutions. These scholars and affiliated institutions are largely the same in the two networks. This is the result of the direct overlapping of scholars serving as panellists in both research assessment, and mainly of the presence in the first panel of members coauthoring papers with members of the second and viceversa. 

\section{Discussion and conclusions}

Research evaluation and especially massive research evaluation such as the British REF have gained a central role in university and research policies, progressively reinforcing their function in gatekeeping, filtering, and legitimating knowledge. Moreover, research evaluation is a type of administrative process and as such it is subjected to specific rules and requirements, among which procedural fairness. Procedural fairness require that the panels of experts managing the evaluation processes have a fair composition. European Peer Review Guide of the European Science Foundation recommend that in the composition of panels ``all affected parties'' are fairly represented. The British REF and the Italian research assessments are formally bound to a fair composition of panels . However, it is not easy to define who the affected parties are and what dimensions to take into account in defining their fair representation. Some characteristics of scholars, such as research field, gender, affiliation and geography are easily verifiable. The analysis of these observable characteristics cannot be sufficient. Cognitive particularism or cronyism can arise if the panel is composed of scholars with a common view of their field, are linked by friendship or have a common view of scientific eminence \citep{cole1979}. This problem possibly arises in the evaluation of a discipline characterized by the coexistence of many schools of thought with different approaches, methodologies and policy recipes. This is precisely the problem that literature has highlighted for economics in the case of the research evaluation in the United Kingdom and in Italy. 

This paper proposes an empirical strategy to test whether, despite the formal adherence to  fair panel composition in terms of easily observable member characteristics such as gender and current university affiliation, it is possible to identify hidden connections between members such that panel composition could be considered unfair. Three main type of connections are defined. The first is the direct or indirect collaboration of panellists in writing articles: a disproportionate diffusion of co-authorships among the members of a panel may indicate that its composition is restricted to people who have personal ties and possibly theoretical or methodological similarities. The second type of connections can be revealed by observing the set of scholarly journals where panellists published: a concentration of panel publication in a relatively restricted set of journals can be considered as an indicator of the lack of intellectual diversity in the panel. The third type of connections can be revealed by looking at whether restricted members have studied in a restricted set of institutions, whether they have had past or current affiliations in a restricted set of institutions, whether they have contributed to a restricted set of magazines, newspapers and blogs. Also in this case the existence of hidden connections can be considered as an indicator of limited intellectual and social diversity in the panel. 

Explorative network analysis is the natural tool for investigating these connections. The main problem with this approach is the defintion of critical thresholds for collaborations, the concentration of publications and the narrowness of affiliated institutions. 

The case study regards the composition of the panels appointed to evaluate research in economics, statistics and business during three Italian research assessment exercises, referring to the years 2004-2010, 2011-2014, 2015-2019. Three panels are then considered. In all the three cases,  candidates to be appointed as panellists responded to a public call. The first two panels were appointed directly by ANVUR board members: the panel members were chosen, in undisclosed proportion, from among the candidates who responded to the public invitation and from outside this group. The third panel, on the other hand, was  selected by lot and exclusively from among those who responded to the public call. Consequently, while the composition of the first two panels is the combined result of the self-selection of the candidates who responded to the call with the panellists' choices by the ANVUR governing board, ANVUR governing board played no role in the composition of the third panel. It is therefore possible to consider the third panel as a control group: if the first two panels appear structurally different from the third in terms of collaborations, concentration of publications and narrowness of affiliated institutions, it can be conjectured that this is due to the intervention of ANVUR governing board in the appointment of the first two panels. Indeed, this research strategy permits to overcome the problem of defining critical thresholds.  

All the evidence obtained suggests that the first two panels are structurally different from the third panel, thus the composition of the first two panels do not appear to have a fair representation in terms of diverse viewpoints, scientific perspectives and scholarly thinking heterogeneity. More specifically, the members of the first two panels had connections in terms of co-authorship, common journals and affiliations much higher than the members of the control group. The first two panels appears filled with a core group of scholars linked by hidden connections, forming a single community with shared intellectual ground, common social relations and even shared policy orientation. As for policy orientation, it is not difficult to recognize in the affiliation networks people and institutions that, according to  \citet{Helgadottir}, shape European policy response to the 2008 recession. 

Three possible objections to this conclusion should be considered. The first one argues that in both the first and the second panel there were still a few non-mainstream scholars, such as Giovanni Dosi. It is possible to counter-argue that the presence of a small minority of members in a quasi-monolithic panel is a form of tokenism, i.e. a symbolic effort to make the panel appear as inclusive and pluralist.

The second objection was argued in an official document \citep{ANVUR2012} in which ANVUR responded to early criticism of the 2004-2010 panel composition.  
The objection is that the panel ``is a group of scholars with a high scientific profile and diversified in terms of skills and geographical origin, in full compliance with the criteria for the selection of panels published on the ANVUR website'' \citep{ANVUR2012}. This objection is valid if the analysis is limited to the \textit{actual}, i.e. at the moment of panel formation, diversities in terms of geography and affiliations. This paper definitively documents that the diversification disappears when considering multiple affiliations, rather than just the the current one, and the academic careers of panel members. Moreover, as noted by \citet[p. 34]{cole1979}, this way of selecting panellists according to their scientific prominence may generate cognitive particularism: scholars who achieve ``eminence tend to favor'' others ``who are similarly situated in the hierarchy of science''.

The third objection is also argued in the official document cited above. It takes seriously the question of the strong co-authorship links among the members of the panel 2004-2010, and it is based on the comparison of the co-authorship network generated by the panel with a control group built by considering the top-20 or top-50 Italian economists in Repec database (www.repec.org). The arguments sounds as follows: the coauthorship network linking the top-50 Italian economists and the president of the panel is similar to the one linking panel members. In Italy, in economics, all the scholars of ``high scientific profile'' collaborate. 
It is possible to counter-argue by reiterating the above argument of the risk of cognitive particularism. Another specific counter-argument is that the choice of the top-20 or top-50 italian economists in Repec is not really a control group, but an \textit{ad hoc} choice. First of all: the call for panellists was open worldwide, hence it is not correct to limit the analysis to Italian economists. Repec does not even represents a complete sample of Italian economists since it ``is based on a limited sample of the research output in Economics and Finance. Only material catalogued in RePEc is considered. {[}..{]} Thus, this list is by no means based on a complete sample'' \citep{repec2017}. Finally, Repec ranking is based on bibliometrics and it is therefore biased towards gender, multidisciplinary methods, and any research orientations pursued by a minority of researchers in their respective disciplines \citep{corsi2019}. 

Once documented that the composition of the first two panels is \textit{unfair}, it follows that also the results of the research assessments should be considered as \textit{unfair}. In other words, procedural unfairness in panel composition determines the general unfairness of the research assessment, and unfair results. The notion of ``unfair'' results should be accurately distinguished from ``biased'' results. If we take seriously the central tenet that a fair panel composition determines not only a just procedure, but also substantively better decisions, some interesting questions arise: to what extent did procedural unfairness result in biased evaluations? How did unfair panel composition result in distortions in the research evaluation results? What were the mechanisms that translated the unfair composition into bias in the evaluations?  What were the effect of excluding from the panels varieties of research experiences?

These questions could be answered by a careful analysis of the micro-data of outcomes of research evaluations. Unluckily, these micro-data, by being generated within administrative processes, are considered confidential and are not disclosed for further independent analyses. In the Italian case, only aggregated results are available. It is therefore impossible to give robust evidence of systematic biases against certain school of thought or methodological approaches. On the basis of partial anonymized micro-data, obtained by appealing to freedom of information act, some works have documented anomalies in the results of the first two research assessment exercises for economics, statistics and business \citep{Baccini_De_Nicolao_2016,Baccini_et_al_2020,Baccini_De_Nicolao_2021}.
In particular \citet{Baccini_De_Nicolao_2021} documented that in the VQR 2004-2010 the panel for economics, statistics and business adopted a protocol for evaluating articles different from the one used in the other research areas. Consequently the agreement between peer-review and bibliometric indicators registered in these areas was anomalously higher than the ones of all the other areas. It can be conjectured that the documented anomalies in results are the outcome of mechanisms such as group thinking that prevent the possibility for panellists to have access to different perspectives and ideas. 

The main lesson to be learned from this work is that the fairness of panel composition is a key issue for the design of fair research evaluation procedures.  The adoption of the prevailing recommendations regarding composition in terms of members' gender, age, affiliation, or research field are important and easily verifiable. However, they are not sufficient to guarantee a fair composition of panels in terms of scholarly thinking, background, or policy orientation. Hence, these recommendations are not necessarily appropriate to prevent the emergence of cognitive particularism in research evaluation. The respect of formal requirements of the compositions in terms of gender, age, and so on does not necessarily prevent an unfair composition due to a careless design of the procedure of appointment: for example, a procedure involving a public call for panellists and then a random draw may not ensure a fair representation of ``all affected parties''. Moreover, it does not prevent an unfair composition due to the capture of the regulator by some segments of the scholarly community. It does not even prevent an unfair composition due to a strategy of the regulator, who is able to predetermine the results of the evaluation by a suitable choice of panellists. 

It is very difficult to draw indications about the rules that  governments, universities, or other institutions should apply for building fair panels in research evaluation. Surely, procedural fairness requires the complete transparency and verifiability of the criteria used for appointing members to a panel. For instance, in case the institution would apply a fair representation of ``all affected parties'', it should clearly define both the population representing ``all affected parties'' and the way in which panellists are chosen from it. In this case, randomization may be an eligible criterion.

The issue of designing fair panels has been predominantly addressed as a problem of fair representation by defining thresholds or quotas for observable characteristics of panelists, such as gender, age, research field. Building panel by respecting these thresholds or quotas poses feasibility problems to the institutions. Adding other criteria and quotas to be met for guaranteeing a fair intellectual composition of panels would make panel appointment practically unfeasible. To this end, it would probably be more effective, though not decisive, to clearly define incompatibilities between members. This may be very easy in relations to some feature emerged in this paper: for instance, neither two co-authors nor a supervisor and a supervisee in the past can be both members of a panel. It is instead much more difficult to define incompatibilities in relation to policy orientations, or methodological and epistemological views. The case study presented here shows that it is possible to test the fairness of panel composition by observing social and intellectual connections among panel members. This test, suitably adapted to the different contexts, can be used by authorities in charge of appointing the panels to verify the fairness of their compositions before evaluation procedures begin. This test, suitably adapted, can be used also by independent observers to challenge the procedural fairness of research evaluations.

\bibliography{references}  

\newpage
\section*{Appendix}\label{Supplementary}

\renewcommand{\thesubsection}{A\arabic{subsection}}
\subsection{The panellists selection process}\label{subsectionA1}

\textbf{2004-2010 panel.} For the VQR 2004-2010, the Ministerial decree established that ANVUR governing board should appoint 450 members divided in 14 area panels, and contemporaneously also the 14 panel presidents. However, the ANVUR governing board acted in disagreement with the Ministerial decree: they appointed firstly the presidents of the panels (list published on 10 October 2011), and after almost two months the panel members (list published on 12 December 2011). (In official documents, ANVUR reported that the appointment of presidents and panel members took place at the same time \citep{benedetto2012}.) The presidents of the panels were consulted during the drafting of the operating rules of the VQR \citep{Anonymous2011}; and it is therefore likely that they had a say in the choice of the other panel members \citep{baccini2016}. 

The members of the panels were chosen largely from a list of scholars realized by the CIVR for the never realized VQR 2004-2008 \citep{ANVUR_2013}. The list was compiled after a public call for experts the deadline of which was 30 June 2010 (https://web.archive.org/web/20100514160659/http://civr.miur.it/modulo.html). ANVUR did not disclose any data on call partecipants. ANVUR described the choice of panel members as a two steps procedure. The first step consisted in defining a set of scholars taking into consideration their qualifications and continuity of scientific production, as well as the evaluation
experience. In the second step, ANVUR had to select panel members by covering all the cultural and research lines within the
areas, by assuring a 20\% of foreign scholars, by having a fair distribution of affiliations and geography, and by paying attention to gender distribution. 
``In a limited number of cases'', but the data were never disclosed, ANVUR chose outside this list. In particular, it selected non-listed names for members of foreign
universities \citep{ANVUR_2013}.

ANVUR presented in the final report data about the distribution of the total set of panellists in terms of gender (23.6\% women), affiliation (Italian or foreign (20\%) affiliation) and geography (Italian scholars are divided according to North, South or Center of Italy) (Table 2.12 in \cite{ANVUR_2013}). According to ANVUR, the data showed that the criteria of fairness defined were respected for the whole set of panellists. No evidence about the respect of the criteria for each single panel was presented.  

For economics, statistics and business area, the panel was composed by 36 members. 6 members were women (16.7\%); 10 members (27.8\%) were affiliated with foreign institutions, but only two did not have Italian first and last name. Tullio Jappelli was selected as the president of the panel.

\textbf{2011-2014 panel.} For the VQR 2011-2014, the selection of panellists started with a public call for experts and ended with the formal approval of the composition of the panels and their presidents by the ANVUR governing board (3 September 2015) \citep{ANVUR2015}. ANVUR finally selected 400 panellists, divided into 16 areas. Also in this case, the selection process proceeded in two steps. The first step of the selection process of panellists was primarily based on ``quality''  ``measured, where possible, by h-index, total number of citations, any awards of scientific merit, analysis of the elements of curriculum vitae in the expression of interest, etc.'' \citep[translation by the authors]{ANVUR2015}. Criteria or thresholds, if any, adopted for this evaluation were not disclosed. In the second step the selection was made by trying to fulfil the following conditions for each panel: coverage of the scientific-discipinary subfield (settore scientifico-disciplinare) inside each area with a number of panellists proportional to the number of expected products to be evaluated; significant percentage of members with foreign affiliation; balanced gender distribution; for Italian candidates, fair distribution of affiliations where possible. If these
criteria could not be met by using the list of candidates who responded to the call, ANVUR could have appointed directly other non-listed scholars. The call for panellists was published on the ANVUR website on 5 May 2015, with a deadline of 5 June 2015, then extended to 15 June 2015. 2,149 candidates (30.4\% women) responded to the call, 171 of whom for Area 13. Area 13 applicants were predominantly from affiliations with Italian institutions (76.4\%)); 30.4\% were women).

The selection process took about two months. The appointed panellists received the official invitation to participate in August 2015. The positive responses were close to 99\%.
Subsequently, ANVUR replaced those who had not accepted the invitation, reaching the final lists. 

For economics, statistics and business area, ANVUR selected 31 members and everybody accepted to participate. One undisclosed member was chosen out of those who responded to the call. 9 members were women (29\%); 7 members (22.5\%) were affiliated with foreign institutions, but only two did not have Italian first and last name. Gabriella Bertocchi was appointed as president.  

\textbf{2015-2019 panel.} The VQR was organized into 17 scientific areas and 1 interdisciplinary area for the evaluation of the activities of the `Third Mission' (the set of activities, beyond teaching and research, with which universities have
direct interaction with society). The area of economics, statistics and business was splitted in two different panels: (13a) ``economics, statistics and business''; and (13b) ``Economics and business sciences''.

For the VQR 2015-2019 the panel selection procedure was different from the previous ones. It started, as the previous ones, with a call for experts. Then the 600 panellists were not chosen by ANVUR governing board, but randomly selected from from those who had applied and met the requirements of high qualification and international experience in research and its evaluation at the time. It should be noted also that, differently from the previous procedures, all the panellists were selected from among the scholars who responded to the call. In particular, applications for being panellist were open from 5 February to 2 March 2020 \citep{ANVUR2020a}. On 11 September 2020, ANVUR published the lists of candidates admitted to the draw \citep{ANVUR2020d}. The list included 4.137 scholars, 350 of whom were candidates for Area 13 panels (174 candidates for panel 13a and 176 for panel 13b). Among these candidates 37.7\% were women; no candidates did have a foreign affiliation. The draw, transmitted also in streaming, took place on 17 September 2020 \citep{ANVUR2020c}. The results of the draw were made available on the same day (https://web.archive.org/web/20220308022222/https://www.anvur.it/attivita/vqr/vqr-2015-2019/gev/esito-del-sorteggio-dei-gev-disciplinari/). The composition of the panels resulted from the draw was later slightly modified by ANVUR governing board, by adding 4 members and by replacing a few of resigning members; these modification were completely disclosed \citep{ANVUR2020b}. To completely exclude ANVUR's intervention in the selection of panel members, we considered the results of the draw for Area ``13a economics, statistics and business'' and Area ``13b Economics and business sciences''  as our ``control group''. In the paper we refer to this group of 40 scholars as the Panel of the  2015-2019. 

Where possible, each panel was formed in compliance with a complex list of requirements: at least 25\% of the members had to be full professors (Professore ordinario); at least 20\%, respectively, had to be associate professor (professore associato) or research fellow (ricercatore) in Italian universities; up
to a maximum of 30\% can be researchers structured at Public Research
Bodies (EPR); at least 5\% had to be researchers in foreign
universities or research bodies. Each panel had to have at least one member for each different recruitment field (settore concorsuale) and for each disciplinary sub-field (settore scientifico-disciplinare) of the area with at least 50 members. The remaining members had to be distributed in proportion to the size of the different fields in the area. Moreover: each gender had to be represented for at least one third; no more than 20\% of the members may have belonged to the panel 2011-2014. Once members of the panels were appointed, the ANVUR governing board identified, choosing among them, the 18 panel coordinators. 

Result of the draw for ``13a economics, statistics and business'' was a list of 22 members; the panel in its final composition was composed by 23 members. Results of the draw for ``13b Economics and business sciences'' was a list of 18 members; the panel in its final composition was composed by 21 members. Overall the two panel included 17 women (38,6\%); no members were affiliated with foreign institutions. The appointed coordinators were respectively Emanuela Marrocu and Maria Rosaria Napolitano.

\begin{figure}[h!]
\setcounter{figure}{0}
\renewcommand{\thefigure}{A\arabic{figure}}
\begin{center}
 \includegraphics[scale=0.8]{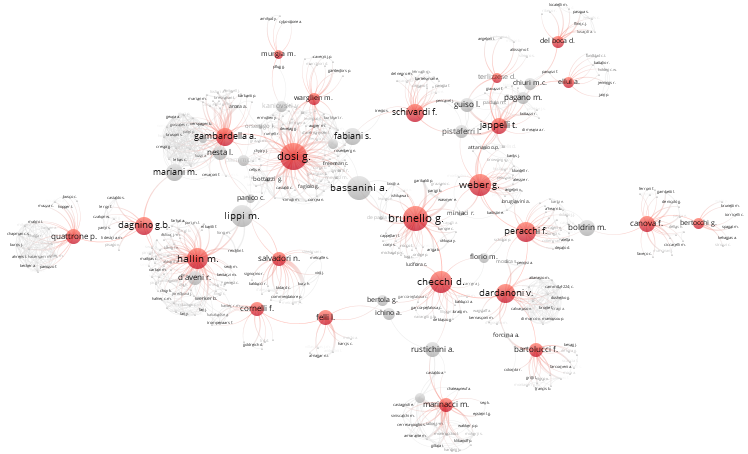}
\caption{The largest component in the co-authorship network for the panel 2004-2010. Size of nodes is proportional to betweenness centrality. Red nodes are panel members.}
\label{figure:2}
  \end{center}
\end{figure}

\begin{figure}[h!]
\renewcommand{\thefigure}{A\arabic{figure}}
\begin{center}
 \includegraphics[scale=0.8]{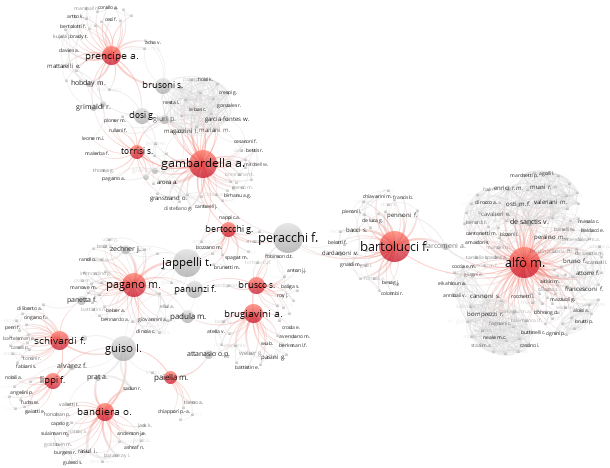}
  \caption{The largest component of the co-authorship network for the panel 2011-2014. Size of vertices is proportional to betweenness centrality. Red nodes are panel members.}
\label{figure:3}
  \end{center}
\end{figure}

\begin{table}[h!]
\setcounter{table}{0}
\renewcommand{\thetable}{A\arabic{table}}
\caption{Betweenness centrality and rank betweenness of the co-authorship network for the panel 2004-2010}
\centering
\label{table:A1}
\begin{tabular}{|p{0.15\linewidth}>{\centering}p{0.09\linewidth}>{\centering}p{0.09\linewidth}>{\centering}p{0.09\linewidth}|p{0.14\linewidth}>{\centering}p{0.09\linewidth}>{\centering}p{0.09\linewidth}>{\centering\arraybackslash}p{0.09\linewidth}|}
\hline
\textbf{Name} & \textbf{Betweenness centrality} & \textbf{Rank Betweenness} & \textbf{panel member} & \textbf{Name} & \textbf{Betweenness centrality} & \textbf{Rank Betweenness} & \textbf{panel member} \\
\hline
Dosi G. & 0.246 & 1 & Yes & D'Aveni R. & 0.023 & 26 & No \\
Brunello G. & 0.183 & 2 & Yes & Marinacci M. & 0.023 & 27 & Yes \\
Bassanini A. & 0.147 & 3 & No & Pistaferri L. & 0.019 & 28 & No \\
Weber G. & 0.116 & 4 & Yes & Pagano M. & 0.017 & 29 & No \\
Checchi D. & 0.116 & 5 & Yes & Ichino A. & 0.017 & 30 & No \\
Lippi M. & 0.115 & 6 & No & Warglien M. & 0.017 & 31 & Yes \\
Hallin M. & 0.096 & 7 & Yes & Chiuri M.C. & 0.016 & 32 & No \\
Peracchi F. & 0.088 & 8 & Yes & Del Boca D. & 0.015 & 33 & Yes \\
Dardanoni V. & 0.066 & 9 & Yes & Ellul A. & 0.013 & 34 & Yes \\
Gambardella A. & 0.064 & 10 & Yes & Bertola G. & 0.013 & 35 & No \\
Dagnino G.B. & 0.058 & 11 & Yes & Pammolli F. & 0.010 & 36 & No \\
Schivardi F. & 0.057 & 12 & Yes & Bertocchi G. & 0.010 & 37 & Yes \\
Mariani M. & 0.045 & 13 & No & Kaniovski Y. & 0.010 & 38 & No \\
Fabiani S. & 0.045 & 14 & No & Terlizzese D. & 0.008 & 39 & Yes \\
Jappelli T. & 0.044 & 15 & Yes & Maccheroni F. & 0.008 & 40 & No \\
Boldrin M. & 0.043 & 16 & No & Marengo L. & 0.008 & 41 & No \\
Canova F. & 0.042 & 17 & Yes & Florio M. & 0.008 & 42 & No \\
Rustichini A. & 0.035 & 18 & No & Guthrie J. & 0.007 & 43 & Yes \\
Quattrone P. & 0.033 & 19 & Yes & Murgia M. & 0.007 & 44 & Yes \\
Felli L. & 0.031 & 20 & Yes & Panico C. & 0.005 & 45 & No \\
Salvadori N. & 0.028 & 21 & Yes & Magazzini L. & 0.004 & 46 & No \\
Bartolucci F. & 0.027 & 22 & Yes & Frino A. & 0.003 & 47 & Yes \\
Cornelli F. & 0.027 & 23 & Yes & Cichelli A. & 0.003 & 48 & Yes \\
Nesta L. & 0.026 & 24 & No & Jones L.R. & 0.002 & 49 & No \\
Guiso L. & 0.024 & 25 & No & Mussari R. & 0.002 & 50 & Yes \\
\hline
\end{tabular}%
\end{table}

\begin{table}[h!]
\renewcommand{\thetable}{A\arabic{table}}
\renewcommand{\thetable}{A\arabic{table}}
\caption{Betweenness centrality and rank betweenness of the co-authorship network for the panel 2011-2014}
\label{table:A2}
\centering
\begin{tabular}{|p{0.15\linewidth} >{\centering}p{0.09\linewidth} >{\centering}p{0.09\linewidth} >{\centering}p{0.09\linewidth} | p{0.14\linewidth} >{\centering}p{0.09\linewidth} >{\centering}p{0.09\linewidth} >{\centering\arraybackslash}p{0.09\linewidth}|}
\hline
\textbf{Name} & \textbf{Betweenness centrality} & \textbf{Rank Betweenness} & \textbf{panel member} & \textbf{Name} & \textbf{Betweenness centrality} & \textbf{Rank Betweenness} & \textbf{panel member} \\
\hline
Bartolucci F. & 0.099 & 1 & Yes & Brusoni S. & 0.012 & 14 & No \\
Alfò M. & 0.097 & 2 & Yes & Dosi G. & 0.011 & 15 & No \\ 
Peracchi F. & 0.096 & 3 & No & Lippi F. & 0.010 & 16 & Yes \\
Jappelli T. & 0.073 & 4 & No & Bertocchi G. & 0.009 & 17 & Yes \\
Gambardella A. & 0.071 & 5 & Yes & Torrisi S. & 0.007 & 18 & Yes \\
Guiso L. & 0.051 & 6 & No & Vivarelli M. & 0.006 & 19 & Yes \\
Pagano M. & 0.041 & 7 & Yes & Greco S. & 0.006 & 20 & Yes \\
Brugiavini A. & 0.019 & 8 & Yes & Padula M. & 0.005 & 21 & No \\
Schivardi F. & 0.019 & 9 & Yes & Paiella M. & 0.005 & 22 & Yes \\
Prencipe A. & 0.018 & 10 & Yes & De Fraja G. & 0.004 & 23 & Yes \\
Bandiera O. & 0.017 & 11 & Yes & Piga C. & 0.003 & 24 & No \\
Panunzi F. & 0.015 & 12 & No & Sarno L. & 0.003 & 25 & Yes\\
Brusco S. & 0.014 & 13 & Yes & Frey M. & 0.002 & 26 & No \\
\hline
\end{tabular}%
\end{table}

\begin{table}[h!]
\renewcommand{\thetable}{A\arabic{table}}
\caption{Betweenness centrality and rank betweenness of the the co-authorship network for the panel 2015-2019}
\label{table:A3}
\centering
\begin{tabular}{|p{0.14\linewidth} >{\centering}p{0.09\linewidth} >{\centering}p{0.09\linewidth} >{\centering}p{0.09\linewidth} | p{0.14\linewidth} >{\centering}p{0.09\linewidth} >{\centering}p{0.09\linewidth} >{\centering\arraybackslash}p{0.09\linewidth}|}
\hline
\textbf{Name} & \textbf{Betweenness centrality} & \textbf{Rank Betweenness} & \textbf{panel member} & \textbf{Name} & \textbf{Betweenness centrality} & \textbf{Rank Betweenness} & \textbf{panel member} \\
\hline
Ruggeri F. & 0.045 & 1 & Yes & Meliciani V. & 0.003 & 9 & Yes \\
Nicolis O. & 0.041 & 2 & No & Mencarini L. & 0.002 & 10 & Yes \\
Stingo F. & 0.036 & 3 & Yes & Pagano M. & 0.002 & 11 & Yes \\
Fassò A. & 0.031 & 4 & Yes & Castellani D. & 0.002 & 12 & No \\
Chiodi M. & 0.016 & 5 & Yes & Piva M. & 0.002 & 13 & Yes \\
Adelfio G. & 0.008 & 6 & Yes & Notarnicola B. & 0.002 & 14 & Yes \\
Bevilacqua M. & 0.007 & 7 & No & Savona M. & 0.002 & 15 & No \\
Perna A. & 0.006 & 8 & Yes & Antonioli D. & 0.002 & 16 & Yes \\
\hline
\end{tabular}%
\end{table}

\begin{table}[h!]
\renewcommand{\thetable}{A\arabic{table}}
\caption{Degree frequency distribution of the journal based network for the panel 2004-2010}
\label{table:9}
\centering
\begin{tabular}{|>{\centering}p{0.08\linewidth}>{\centering}p{0.08\linewidth}>{\centering}p{0.09\linewidth}|>{\centering}p{0.08\linewidth}>{\centering}p{0.08\linewidth}>{\centering}p{0.09\linewidth}|>{\centering}p{0.08\linewidth}>{\centering}p{0.08\linewidth}>{\centering\arraybackslash}p{0.09\linewidth}|}
\hline
\textbf{Degree} & \textbf{Frequency} & \textbf{Frequency (\%)} &
\textbf{Degree} & \textbf{Frequency} & \textbf{Frequency (\%)} &
\textbf{Degree} & \textbf{Frequency} & \textbf{Frequency (\%)} \\
\hline
0 & 2 & 5.56 & 6 & 4 & 11.11 & 14 & 1 & 2.78 \\
1 & 1 & 2.78 & 7 & 2 & 5.56 & 15 & 1 & 2.78 \\
2 & 2 & 5.56 & 9 & 2 & 5.56 & 16 & 4 & 11.11 \\
3 & 3 & 8.33 & 11 & 2 & 5.56 & 17 & 2 & 5.56 \\
4 & 2 & 5.56 & 12 & 3 & 8.33 & & & \\
5 & 4 & 11.11 & 13 & 1 & 2.78 & & & \\
\hline
\end{tabular}%
\end{table}

\begin{table}[h!]
\renewcommand{\thetable}{A\arabic{table}}
\caption{Degree frequency distribution of the journal based network for the panel 2011-2014}
\label{table:11}
\centering
\begin{tabular}{|>{\centering}p{0.08\linewidth}>{\centering}p{0.08\linewidth}>{\centering}p{0.09\linewidth}|>{\centering}p{0.08\linewidth}>{\centering}p{0.08\linewidth}>{\centering}p{0.09\linewidth}|>{\centering}p{0.08\linewidth}>{\centering}p{0.08\linewidth}>{\centering\arraybackslash}p{0.09\linewidth}|}
\hline
\textbf{Degree} & \textbf{Frequency} & \textbf{Frequency (\%)} &
\textbf{Degree} & \textbf{Frequency} & \textbf{Frequency (\%)} &
\textbf{Degree} & \textbf{Frequency} & \textbf{Frequency (\%)} \\
\hline
0 & 3 & 9.68 & 6 & 2 & 6.45 & 14 & 5 & 16.13 \\
1 & 2 & 6.45 & 7 & 1 & 3.23 & 15 & 1 & 3.23 \\
2 & 2 & 6.45 & 8 & 2 & 6.45 & 17 & 2 & 6.45 \\
3 & 1 & 3.23 & 10 & 2 & 6.45 & 19 & 1 & 3.23 \\
4 & 1 & 3.23 & 12 & 2 & 6.45 & & & \\
5 & 1 & 3.23 & 13 & 3 & 9.68 & & & \\
\hline
\end{tabular}%
\end{table}

\begin{table}[h!]
\renewcommand{\thetable}{A\arabic{table}}
\caption{Degree frequency distribution of the journal based network for the panel 2015-2019}
\label{table:13}
\centering
\begin{tabular}{|>{\centering}p{0.08\linewidth}>{\centering}p{0.08\linewidth}>{\centering}p{0.09\linewidth}|>{\centering}p{0.08\linewidth}>{\centering}p{0.08\linewidth}>{\centering\arraybackslash}p{0.09\linewidth}|}
\hline
\textbf{Degree} & \textbf{Frequency} & \textbf{Frequency (\%)} &
\textbf{Degree} & \textbf{Frequency} & \textbf{Frequency (\%)} \\
\hline
1 & 3 & 8.11 & 9 & 1 & 2.70 \\
3 & 3 & 8.11 & 10 & 3 & 8.11 \\
4 & 2 & 5.41 & 11 & 2 & 5.41 \\
5 & 5 & 13.51 & 12 & 1 & 2.70 \\
6 & 3 & 8.11 & 13 & 2 & 5.41 \\
7 & 4 & 10.81 & 14 & 1 & 2.70 \\
8 & 6 & 16.22 & 22 & 1 & 2.70 \\
\hline
\end{tabular}%
\end{table}


\begin{table}[h!]
\renewcommand{\thetable}{A\arabic{table}}
\caption{Betweenness centrality, rank betweenness and island of the journal based network of scholars of the panel 2004-2010}
\label{table:10}
\centering
\begin{tabular}{|p{0.15\linewidth} >{\centering}p{0.09\linewidth} >{\centering}p{0.09\linewidth} >{\centering}p{0.09\linewidth} | p{0.14\linewidth} >{\centering}p{0.09\linewidth} >{\centering}p{0.09\linewidth} >{\centering\arraybackslash}p{0.09\linewidth}|}
\hline
\textbf{Name} & \textbf{Betweenness centrality} & \textbf{Rank Betweenness} & \textbf{Island Color} & \textbf{Name} & \textbf{Betweenness centrality} & \textbf{Rank Betweenness} & \textbf{Island Color} \\
\hline
Dosi G. & 0.138 & 1 & Blue & Ellul A. & 0.018 & 19 & Green \\
Frino A. & 0.125 & 2 & Grey & Bottazzi L. & 0.015 & 20 & Green \\
Guthrie J. & 0.081 & 3 & Red & Frittelli M. & 0.010 & 21 & Grey \\
Bisin A. & 0.079 & 4 & Green & Marinacci M. & 0.010 & 22 & Green \\
Jappelli T. & 0.058 & 5 & Green & Hallin M. & 0.006 & 23 & Green \\
Schivardi F. & 0.051 & 6 & Green & Checchi D. & 0.006 & 24 & Green \\
Brunello G. & 0.050 & 7 & Green & Quattrone P. & 0.005 & 25 & Red \\
Peracchi F. & 0.050 & 8 & Green & Bartolucci F. & 0.004 & 26 & Green \\
Gambardella A. & 0.048 & 9 & Blue & Salvadori N. & 0.001 & 27 & Grey \\
Weber G. & 0.042 & 10 & Green & Cornelli F. & 0.001 & 28 & Green \\
Dagnino G.B. & 0.041 & 11 & Grey & Del Boca D. & 0.001 & 29 & Green \\
Bertocchi G. & 0.031 & 12 & Green & Bergami M. & 0.000 & 30 & Grey \\
Rossi B. & 0.031 & 13 & Green & Cichelli A. & 0.000 & 31 & Grey \\
Warglien M. & 0.028 & 14 & Blue & Guido G. & 0.000 & 32 & Grey \\
Canova F. & 0.025 & 15 & Green & Murgia M. & 0.000 & 33 & Grey \\
Dardanoni V. & 0.022 & 16 & Green & Mussari R. & 0.000 & 34 & Red \\
Felli L. & 0.020 & 17 & Green & Terlizzese D. & 0.000 & 36 & Green \\
Ronchetti E. & 0.020 & 18 & Green & Zamagni V. & 0.000 & 35 & Grey \\
\hline
\end{tabular}%
\end{table}

\begin{table}[h!]
\renewcommand{\thetable}{A\arabic{table}}
\caption{Betweenness centrality, rank betweenness and island of the journal based network of scholars of the panel 2011-2014}
\label{table:12}
\centering
\begin{tabular}{|p{0.15\linewidth} >{\centering}p{0.09\linewidth} >{\centering}p{0.09\linewidth} >{\centering}p{0.09\linewidth} | p{0.14\linewidth} >{\centering}p{0.09\linewidth} >{\centering}p{0.09\linewidth} >{\centering\arraybackslash}p{0.09\linewidth}|}
\hline
\textbf{Name} & \textbf{Betweenness centrality} & \textbf{Rank Betweenness} & \textbf{Island Color} & \textbf{Name} & \textbf{Betweenness centrality} & \textbf{Rank Betweenness} & \textbf{Island Color} \\
\hline
Gambardella A. & 0.130 & 1 & Green & Brusco S. & 0.009 & 17 & Green \\
Cinquini L. & 0.115 & 2 & Grey & Alfò M. & 0.009 & 18 & Red \\
Sarno L. & 0.100 & 3 & Green & Paiella M. & 0.008 & 19 & Green \\
Fiordelisi F. & 0.076 & 4 & Grey & Corbetta G. & 0.008 & 20 & Grey \\
Vivarelli M. & 0.060 & 5 & Green & Brugiavini A. & 0.006 & 21 & Green \\
Larrinaga-G. C. & 0.060 & 6 & Grey & Ronchetti E. & 0.003 & 22 & Red \\
Bertocchi G. & 0.052 & 7 & Green & Bartolucci F. & 0.003 & 23 & Red \\
Prencipe A. & 0.043 & 8 & Green & Torrisi S. & 0.002 & 24 & Green \\
Schivardi F. & 0.040 & 9 & Green & Asso P.F. & 0.002 & 25 & Grey \\
Bandiera O. & 0.027 & 10 & Green & De Carlo M. & 0.000 & 26 & Grey \\
Pagano M. & 0.025 & 11 & Green &  De Paola M. & 0.000 & 27 & Grey \\
Cavaliere G. & 0.023 & 12 & Green &  Greco S. & 0.000 & 28 & Grey \\
De Fraja G. & 0.023 & 13 & Green & Malanima P. & 0.000 & 29 & Grey \\
Kretschmer T. & 0.021 & 14 & Green &  Rosazza G. E. & 0.000 & 30 & Grey \\
Pesavento E. & 0.016 & 15 & Green & Salomone R. & 0.000 & 31 & Grey \\
Lippi F. & 0.011 & 16 & Green & & & & \\
\hline
\end{tabular}%
\end{table}

\begin{table}[h!]
\renewcommand{\thetable}{A\arabic{table}}
\caption{Betweenness centrality, rank betweenness and island of the journal based network of scholars of the panel 2015-2019}
\label{table:14}
\centering
\begin{tabular}{|p{0.16\linewidth} >{\centering}p{0.09\linewidth} >{\centering}p{0.09\linewidth} >{\centering}p{0.09\linewidth} | p{0.14\linewidth} >{\centering}p{0.09\linewidth} >{\centering}p{0.09\linewidth} >{\centering\arraybackslash}p{0.09\linewidth}|}
\hline
\textbf{Name} & \textbf{Betweenness centrality} & \textbf{Rank Betweenness} & \textbf{Island Color} & \textbf{Name} & \textbf{Betweenness centrality} & \textbf{Rank Betweenness} & \textbf{Island Color} \\
\hline
Piva M. & 0.272 & 1 & Green & Cavallo B. & 0.016 & 20 & Grey \\
Michetti E. & 0.063 & 2 & Grey & Adelfio G. & 0.010 & 21 & Blue \\
Marrocu E. & 0.061 & 3 & Green & La Rosa F. & 0.009 & 22 & Red \\
Pretaroli R. & 0.059 & 4 & Grey & Bracci E. & 0.008 & 23 & Red \\
Berni R. & 0.059 & 5 & Blue & Corsi K. & 0.008 & 24 & Red \\
Ruggeri F. & 0.059 & 6 & Blue & Antonioli D. & 0.007 & 25 & Green \\
Pagano M. & 0.056 & 7 & Grey & Fassò A. & 0.006 & 26 & Blue \\
Opocher A. & 0.056 & 8 & Grey & Mavilia R. & 0.004 & 27 & Green \\
Vasta M. & 0.056 & 9 & Grey & Stingo F. & 0.004 & 28 & Blue \\
Bajo E. & 0.054 & 10 & Grey & Chiodi M. & 0.004 & 29 & Blue \\
Picchio M. & 0.047 & 11 & Grey & Levrero E.S. & 0.004 & 30 & Grey \\
Gaeta G.L. & 0.046 & 12 & Grey & Scalzo V. & 0.003 & 31 & Grey \\
Meliciani V. & 0.041 & 13 & Green & Avallone P. & 0.000 & 32 & Grey \\
Perna A. & 0.038 & 14 & Grey & Dell'Atti S. & 0.000 & 33 & Grey \\
Napolitano M.R. & 0.034 & 15 & Grey & Ghellini G. & 0.000 & 34 & Grey \\
Altomonte C. & 0.034 & 16 & Green & Guerriero C. & 0.000 & 35 & Grey \\
Mencarini L. & 0.027 & 17 & Grey & Mason M. & 0.000 & 36 & Grey \\
Greco G. & 0.020 & 18 & Grey & Mosca M. & 0.000 & 37 & Grey \\
Notarnicola B. & 0.019 & 19 & Grey & & & & \\
\hline
\end{tabular}%
\end{table}

\begin{table}[h!]
\renewcommand{\thetable}{A\arabic{table}}
\caption{The important vertices for the affinity network of the panel 2004-2010. The number of important vertices is fixed to 17, about a half of the size of the panel.}
\label{table:17}
\centering
\begin{tabular}{|m{0.15\linewidth} >{\centering}m{0.07\linewidth} >{\centering}m{0.07\linewidth} >{\centering}m{0.07\linewidth}>{\centering}m{0.07\linewidth} | m{0.13\linewidth} >{\centering}m{0.07\linewidth} >{\centering}m{0.07\linewidth} >{\centering\arraybackslash}m{0.07\linewidth}|}
\hline
\multicolumn{5}{|c|}{\textbf{Scholars}} & \multicolumn{4}{|c|}{\textbf{Affiliated institutions}}\\
\hline
\textbf{Name} & \textbf{\makecell{Eigenve- \\ ctor cen- \\ trality}} & \textbf{Rank Eigenvector} & \textbf{Island Color} & \textbf{Panel member} & \textbf{Name} & \textbf{\makecell{Eigenve- \\ ctor cen- \\ trality}} & \textbf{Rank Eigenvector} & \textbf{Island Color} \\
\hline
Pagano M. & 0.307 & 1 & Green & No & CEPR & 0.514 & 1 & Green \\
Ichino A. & 0.286 & 2 & Green & No & NBER & 0.394 & 2 & Green \\
Schivardi F. & 0.255 & 3 & Green & Yes & Bocconi U. & 0.345 & 3 & Green \\
Guiso L. & 0.252 & 4 & Green & No & EIEF & 0.289 & 4 & Green \\
Pistaferri L. & 0.239 & 5 & Green & No & lavoce.info & 0.275 & 5 & Green \\
Bisin A. & 0.234 & 6 & Green & Yes & IGIER & 0.196 & 6 & Green \\
Bottazzi L. & 0.227 & 7 & Green & Yes & IlSole24Ore & 0.176 & 7 & Green \\
Rustichini A. & 0.224 & 8 & Green & No & LSE & 0.153 & 8 & Green \\
Checchi D. & 0.214 & 9 & Green & Yes & MIT & 0.148 & 9 & Green \\
Peracchi F. & 0.203 & 10 & Green & Yes & IZA & 0.126 & 10 & Green \\
Ellul A. & 0.189 & 11 & Green & Yes & Bank of Italy & 0.114 & 11 & Green \\
Rossi B. & 0.183 & 12 & Grey & Yes & EEA & 0.110 & 12 & Green \\
Boldrin M. & 0.183 & 13 & Green & No & European UI & 0.105 & 13 & Green \\
Terlizzese D. & 0.173 & 14 & Green & Yes & Il Foglio & 0.101 & 14 & Green \\
Weber G. & 0.168 & 15 & Green & Yes & Bologna U. & 0.094 & 15 & Green \\
Gambardella A. & 0.166 & 16 & Grey & Yes & Cambridge U. & 0.085 & 16 & Grey \\
Jappelli T. & 0.157 & 17 & Green & Yes & CSEF & 0.076 & 17 & Grey \\
\hline
\end{tabular}%
\end{table}

\begin{table}[h!]
\renewcommand{\thetable}{A\arabic{table}}
\caption{The important vertices for the affinity network of the panel 2011-2014. The number of important vertices is fixed to 17, about a half of the size of the panel.}
\label{table:19}
\centering
\begin{tabular}{|m{0.15\linewidth} >{\centering}m{0.07\linewidth} >{\centering}m{0.07\linewidth} >{\centering}m{0.07\linewidth}>{\centering}m{0.07\linewidth} | m{0.20\linewidth} >{\centering}m{0.07\linewidth} >{\centering}m{0.07\linewidth} >{\centering\arraybackslash}m{0.07\linewidth}|}
\hline
\multicolumn{5}{|c|}{\textbf{Network of scholars}} & \multicolumn{4}{|c|}{\textbf{Network of affiliation}}\\
\hline
\textbf{Name} & \textbf{\makecell{Eigenve- \\ ctor cen- \\ trality}} & \textbf{Rank Eigenvector} & \textbf{Island Color} & \textbf{panel member} & \textbf{Name} & \textbf{\makecell{Eigenve- \\ ctor cen- \\ trality}} & \textbf{Rank Eigenvector} & \textbf{Island Color} \\
\hline
Pagano M. & 0.350 & 1 & Green & Yes & CEPR & 0.519 & 1 & Green \\
Schivardi F. & 0.294 & 2 & Green & Yes & NBER & 0.383 & 2 & Green \\
Guiso L. & 0.287 & 3 & Green & No & lavoce.info & 0.348 & 3 & Green \\
Padula M. & 0.285 & 4 & Green & No & Bocconi U. & 0.341 & 4 & Green \\
Panunzi F. & 0.282 & 5 & Green & No & EIEF & 0.272 & 5 & Green \\
Lippi F. & 0.239 & 6 & Green & Yes & Bank of Italy & 0.136 & 6 & Green \\
Brugiavini A. & 0.216 & 7 & Green & Yes & IGIER & 0.119 & 7 & Green \\
Peracchi F. & 0.215 & 8 & Green & No & LSE & 0.116 & 8 & Green \\
Jappelli T. & 0.214 & 9 & Green & No & Roma Tor Vergata U. & 0.112 & 9 &
Grey \\
Bandiera O. & 0.199 & 10 & Green & Yes & Napoli Federico II U. & 0.110 & 10
& Grey \\
Brusco S. & 0.196 & 11 & Green & Yes & CSEF & 0.108 & 11 & Grey \\
Torrisi S. & 0.196 & 12 & Green & Yes & CFS & 0.108 & 12 & Grey \\
Gambardella A. & 0.195 & 13 & Green & Yes & ECGI & 0.108 & 13 & Grey \\
Sarno L. & 0.193 & 14 & Green & Yes & ECB & 0.100 & 14 & Grey \\
Paiella M. & 0.166 & 15 & Grey & Yes & EEA & 0.094 & 15 & Grey \\
De Paola M. & 0.148 & 16 & Grey & Yes & IlSole24Ore & 0.093 & 16 & Grey \\
Bertocchi G. & 0.144 & 17 & Grey & Yes & Il Foglio & 0.091 & 17 & Grey \\
\hline
\end{tabular}%
\end{table}

\begin{table}[h!]
\renewcommand{\thetable}{A\arabic{table}}
\caption{The important vertices for the affinity network of the panel 2015-2019. The number of important vertices is fixed to 17, about a half of the size of the panel.}
\label{table:21}
\centering
\begin{tabular}{|m{0.14\linewidth} >{\centering}m{0.07\linewidth} >{\centering}m{0.07\linewidth} >{\centering}m{0.09\linewidth}>{\centering}m{0.07\linewidth} | m{0.20\linewidth} >{\centering}m{0.07\linewidth} >{\centering}m{0.07\linewidth} >{\centering\arraybackslash}m{0.07\linewidth}|}
\hline
\multicolumn{5}{|c|}{\textbf{Network of scholars}} & \multicolumn{4}{|c|}{\textbf{Network of affiliation}}\\
\hline
\textbf{Name} & \textbf{\makecell{Eigenve- \\ ctor cen- \\ trality}} & \textbf{Rank Eigenvector} & \textbf{Island Color} & \textbf{Panel member} & \textbf{Name} & \textbf{\makecell{Eigenve- \\ ctor cen- \\ trality}} & \textbf{Rank Eigenvector} & \textbf{Island Color} \\
\hline 
Pagano M. & 0.471 & 1 & Green & Yes & Bocconi U. & 0.364 & 1 & Green \\
Mencarini L. & 0.415 & 2 & Grey & Yes & lavoce.info & 0.341 & 2 & Green \\
Altomonte C. & 0.385 & 3 & Green & Yes & SIS & 0.283 & 3 & Violet \\
Guerriero C. & 0.312 & 4 & Green & Yes & CEPR & 0.252 & 4 & Green\\
Picchio M. & 0.247 & 5 & Green & Yes & Firenze U. & 0.216 & 5 & Violet \\
Nicolis O. & 0.203 & 6 & Violet & No & ECB & 0.176 & 6  & Green \\
Mavilia R. & 0.185 & 7 & Grey & Yes & IlSole24Ore & 0.176 & 7  & Green \\
Fassò A. & 0.158 & 8 & Violet & Yes & Bologna U. & 0.175 & 8  & Brown \\
Vasta M. & 0.144 & 9 & Grey & Yes & Cambridge U. & 0.161 & 9  & Green \\
Bajo E. & 0.143 & 10 & Grey & Yes & EIEF & 0.161 & 10 & Green \\
Savona M.& 0.137 & 11 & Yellow & No & Torino U. & 0.153 & 11 & Violet \\
Mosca M. & 0.133 & 12 & Light Blue & Yes & CCA & 0.150 & 12 & Green \\
Chiodi M. & 0.125 & 13 & Violet & Yes & Napoli Federico II U. & 0.135 & 13 & Orange \\
Ghellini G. & 0.124 & 14 & Grey & Yes & UCL - Louvain & 0.130 & 14 & Green \\
Adelfio G. & 0.121 & 15 & Violet & Yes & ISPI & 0.117 & 15 & Green \\
Berni R. & 0.119 & 16 & Grey & Yes & Padova U. & 0.111 & 16 & Violet \\
Stingo F. & 0.113 & 17 & Grey & Yes & GLO & 0.105 & 17 & Green \\
\hline
\end{tabular}%
\end{table}

\begin{table}[h!]
\renewcommand{\thetable}{A\arabic{table}}
\centering
\caption{Frequency distribution of island values of panel 2004-2010 affinity network}
\label{table:18}
\begin{tabular}{|>{\centering}m{0.1\linewidth} >{\centering}m{0.1\linewidth} >{\centering}m{0.1\linewidth} >{\centering}m{0.1\linewidth} | >{\centering}m{0.1\linewidth} >{\centering}m{0.1\linewidth} >{\centering}m{0.1\linewidth} >{\centering\arraybackslash}m{0.1\linewidth}|}
\hline
\multicolumn{4}{|c|}{\textbf{Network of scholars}} & \multicolumn{4}{|c|}{\textbf{Network of affiliation}}\\
\hline
\textbf{Island} & \textbf{Scholars Number} & \textbf{Important vertices Freq} & \textbf{Important vertices Freq\%} & \textbf{Island} & \textbf{Affiliations Number} & \textbf{Important vertices Freq} & \textbf{Important vertices Freq\%} \\
\hline
Grey & 36 & 2 & 11.8 & Grey & 153 & 2 & 11.8 \\
Red & 2 & 0 & 0 & Red & 5 & 0 & 0 \\
Yellow & 2 & 0 & 0 & Yellow & 2 & 0 & 0 \\
Blue & 2 & 0 & 0 & Light Blue & 6 & 0 & 0 \\
Green & 16 & 15 & 88.2 & Green & 15 & 15 & 88.2  \\
& & & & Violet & 2 & 0 & 0 \\
& & & & Pink & 4 & 0 & 0  \\
& & & & Orange & 2 & 0 & 0 \\
& & & & Brown & 2 & 0 & 0 \\
\hline
\textbf{Sum} & 58 & 17 & 100 & \textbf{Sum} & 191 & 17 & 100 \\
\hline
\end{tabular}%
\end{table}

\begin{table}[h!]
\renewcommand{\thetable}{A\arabic{table}}
\caption{Frequency distribution of island values of panel 2011-2014 affinity network}
\label{table:20}
\centering
\begin{tabular}{|>{\centering}m{0.1\linewidth} >{\centering}m{0.1\linewidth} >{\centering}m{0.1\linewidth} >{\centering}m{0.1\linewidth} | >{\centering}m{0.1\linewidth} >{\centering}m{0.1\linewidth} >{\centering}m{0.1\linewidth} >{\centering\arraybackslash}m{0.1\linewidth}|}
\hline
\multicolumn{4}{|c|}{\textbf{Network of scholars}} & \multicolumn{4}{|c|}{\textbf{Network of affiliation}}\\
\hline
\textbf{Island} & \textbf{Scholars Number} & \textbf{Important vertices Freq} & \textbf{Important vertices Freq\%} & \textbf{Island} & \textbf{Affiliations Number} & \textbf{Important vertices Freq} & \textbf{Important vertices Freq\%} \\
\hline
Grey & 23 & 3 & 17.7 & Grey & 131 & 9 & 52.9 \\
Red & 2 & 0 & 0 & Red & 8 & 0 & 0 \\
Green & 15 & 14 & 82.3 & Green & 8 & 8 & 47.1 \\
\hline
\textbf{Sum} & 40 & 17 & 100 & \textbf{Sum} & 147 & 17 & 100 \\
\hline
\end{tabular}%
\end{table}

\begin{table}[h!]
\renewcommand{\thetable}{A\arabic{table}}
\caption{Frequency distribution of island values of panel 2015-2019 affinity network}
\label{table:22}
\centering
\begin{tabular}{|>{\centering}m{0.1\linewidth} >{\centering}m{0.1\linewidth} >{\centering}m{0.1\linewidth} >{\centering}m{0.1\linewidth} | >{\centering}m{0.1\linewidth} >{\centering}m{0.1\linewidth} >{\centering}m{0.1\linewidth} >{\centering\arraybackslash}m{0.1\linewidth}|}
\hline
\multicolumn{4}{|c|}{\textbf{Network of scholars}} & \multicolumn{4}{|c|}{\textbf{Network of affiliation}}\\
\hline
\textbf{Island Color} & \textbf{Scholars Number} & \textbf{Important vertices Freq} & \textbf{Important vertices Freq\%} & \textbf{Island Color} & \textbf{Affiliations Number} & \textbf{Important vertices Freq} & \textbf{Important vertices Freq\%} \\
\hline
Grey & 22 & 7 & 41.2 & Grey & 131 & 0 & 0 \\
Blue & 4 & 0 & 0 & Blue & 3 & 0 & 0 \\
Brown & 2 & 0 & 0 & Brown & 2 & 1 & 5.9 \\
Orange & 3 & 0 & 0 & Orange & 3 & 1 & 5.9 \\
Light Blue & 2 & 1 & 5.9 & Light Blue & 2 & 0 & 0  \\
Violet & 4 & 4 & 23.5 & Violet & 9 & 4 & 23.5 \\
Green & 4 & 4 & 23.5 & Green & 15 & 11 & 64.7  \\
Yellow & 3 & 1 & 5.9 & Yellow & 2 & 0 & 0 \\
& & & &  Red & 4 & 0 & 0 \\
\hline
\textbf{Sum} & 44 & 17 & 100 & \textbf{Sum} & 171 & 17 & 100 \\
\hline
\end{tabular}%
\end{table}

\end{document}